\documentclass[10pt,english,journal]{IEEEtran}
\usepackage[T1]{fontenc}
\usepackage[latin9]{inputenc}
\usepackage{color}
\usepackage{babel}
\usepackage{amsmath}
\usepackage{amssymb}
\usepackage{graphicx}
\PassOptionsToPackage{normalem}{ulem}
\usepackage{ulem}
\usepackage[unicode=true,
 bookmarks=true,bookmarksnumbered=true,bookmarksopen=true,bookmarksopenlevel=1,
 breaklinks=false,pdfborder={0 0 0},pdfborderstyle={},backref=false,colorlinks=false]
 {hyperref}
\hypersetup{pdftitle={Your Title},
 pdfauthor={Your Name},
 pdfpagelayout=OneColumn, pdfnewwindow=true, pdfstartview=XYZ, plainpages=false}

\makeatletter

\providecommand{\tabularnewline}{\\}

\usepackage[caption=false,font=footnotesize]{subfig}
\usepackage[normalem]{ulem}

\@ifundefined{showcaptionsetup}{}{%
 \PassOptionsToPackage{caption=false}{subfig}}
\usepackage{subfig}
\makeatother

\begin{document}

\title{Hybrid Beamforming Based on Implicit Channel State Information for
Millimeter Wave Links }

\author{Hsiao-Lan~Chiang, Wolfgang Rave, Tobias Kadur, and Gerhard Fettweis,
\IEEEmembership{Fellow, IEEE}\thanks{Hsiao-Lan Chiang, Wolfgang Rave, Tobias Kadur, and Gerhard Fettweis
are with the Vodafone Chair Mobile Communications Systems, Technische
Universität Dresden, Germany, e-mail: \protect\protect\href{http://Hsiao-lan.Chiang, Wolfgang.Rave, Tobias.Kadur, Gerhard.Fettweis@tu-dresden.de}{Hsiao-lan.Chiang, Wolfgang.Rave, Tobias.Kadur, Gerhard.Fettweis@tu-dresden.de}.}\thanks{The research leading to these results has received funding from the
European Union\textquoteright s Horizon 2020 research and innovation
programme under grant agreement No. 671551 (5G-XHaul) and the TUD-NEC
project \textquotedblleft mmWave Antenna Array Concept Study\textquotedblright ,
a cooperation project between Technische Universität Dresden (TUD),
Germany, and NEC, Japan.}\thanks{This work has been presented in part at the IEEE International Communications
Conference (ICC), Kansas City, MO, USA, May 2018.}}
\maketitle
\begin{abstract}
Hybrid beamforming provides a promising solution to achieve high data
rate transmission at millimeter waves. Implementing hybrid beamforming
at a transceiver based on available channel state information is a
common solution. However, many reference methods ignore the complexity
of channel estimation for large antenna arrays or subsequent steps,
such as the singular value decomposition of a channel matrix. To this
end, we present a low-complexity scheme that exploits implicit channel
knowledge to facilitate the design of hybrid beamforming for frequency-selective
fading channels. The implicit channel knowledge can be interpreted
as couplings between all possible pairs of analog beamforming vectors
at the transmitter and receiver over the surrounding channel. Instead
of calculating mutual information between large antenna arrays, we
focus on small-size coupling matrices between beam patterns selected
by using appropriate key parameters as performance indicators. This
converts the complicated hybrid beamforming problem to a much simpler
one: it amounts to collecting different sets of the large-power coupling
coefficients to construct multiple alternatives for an effective channel
matrix. Then, the set yielding the largest Frobenius norm (or the
largest absolute value of the determinant) of the effective channel
provides the solution to the hybrid beamforming problem. It turns
out that the proposed method does not require information on MIMO
channel and can be simply implemented by the received correlated pilot
signals that are supposed to be used for channel estimation.
\end{abstract}


\section{Introduction}

\IEEEPARstart{W}{ith} the rapid increase of data rates in wireless
communications, the problem of bandwidth shortage is getting more
critical. Therefore, there is a growing interest in using millimeter
wave (mmWave) for future wireless communications, taking advantage
of an enormous amount of available spectrum at frequencies $>6$ GHz
\cite{Rappaport2014}. Measurements of mmWave channel characteristics
presented in \cite{Thomas2014}-\nocite{Rappaport2015}\nocite{Haneda2016}\cite{3GPP38900}
show that the path loss in such an environment is very severe. In
order to improve capacity and service quality, mmWave small-cell deployment
together with beamforming for large antenna arrays is seen as a promising
approach \cite{Rappaport2013,Frascolla2014}. When a system operates
at mmWave frequency bands, it is infeasible to equip each antenna
with its own radio frequency (RF) chain due to high implementation
cost and power consumption. Accordingly, a combination of analog beamforming
(operating in passband) \cite{Liberti1999,Hajimiri2005} and digital
beamforming (operating in baseband) \cite{VanTrees2004} can be one
of the low-cost solutions, and this combination is commonly called
hybrid beamforming \cite{Zhang2005}-\nocite{Ayach2014}\nocite{Roh2014}\cite{Han2015}.

In hybrid beamforming systems, although both analog and digital beamforming
matrices use the same word \textit{beamforming}, only the former has
a specific geometrical meaning in the sense of directing or collecting
energy towards specific directions by using antenna arrays. In contrast,
the digital beamforming matrix has rather an algebraic meaning. In
other words, it is more like a coefficient matrix. \textcolor{black}{According
to the functions of analog and digital beamforming,} hybrid beamforming
can be regarded as first converting an over-the-air large-scale MIMO
channel matrix $\mathbf{H}$ in the spatial domain into an effective
channel $\mathbf{H}_{E}$ of significantly smaller size\footnote{The size of $\mathbf{H}_{E}$ is determined by the number of available
RF chains on both sides; a more detailed description of this will
be given in Section \ref{sec:System_model}.} in the angular domain by analog beamforming vectors. Then, one can
further try linear combinations of analog beamforming vectors with
entries of digital beamforming matrices as coefficients to maximize
mutual information conditional on $\mathbf{H}_{E}$.

Unquestionably, it is intractable to deal with hybrid beamforming
at a transmitter and a receiver simultaneously. To simplify the
problem, one can assume that the channel state information (CSI) is
available. Then, by utilizing the singular value decomposition (SVD)
of the channel matrix, the problem of hybrid beamforming on both sides
(i.e., finding the precoder and combiner) can be decoupled and formulated
as two minimization problems \cite{Ayach2014,Ayach2012}\nocite{Alkhateeb2014}-\nocite{Mendez-Rial2016}\cite{Chiang2016_PIMRC}.
In addition, the properties of codebooks used for the analog beamforming
are incorporated into the problem as an additional constraint. For
frequency-flat fading channels, a single SVD computation suffices,
while more than one becomes necessary to handle a multi-carrier modulation
in frequency-selective fading channels. One can also decouple the
transceiver by an assumption that either the transmitter or receiver
employs fully digital beamforming to facilitate the problem-solving
process \cite{Alkhateeb2016b}-\nocite{Sohrabi2016b}\cite{Sohrabi2016}.
Nevertheless, most previously proposed hybrid beamforming methods
require channel knowledge and ignore the overhead of channel estimation
for large-scale antenna arrays \cite{Chiang2016_ISWCS}-\nocite{Gao2016}\nocite{Xiao2017}\cite{Venugopal2017}. 

A feasible alternative that decouples the precoder and combiner
is therefore introduced as follows. Generally speaking, given candidates
for the analog beamforming vectors selected from codebooks on both
sides, finding the corresponding optimal digital beamforming is trivial
in the sense that it is almost equivalent to the conventional fully
digital beamforming apart from different power constraints \cite{Telatar1999}.
As a result, the critical issue of hybrid beamforming is definitely
in analog beam selection. The work in \cite{Chiang2017_ICASSP} explains
why analog beam selection based on the power of received correlated
pilot signals is equivalent to the selection method by the orthogonal
matching pursuit (OMP) algorithm \cite{Cai2011}. It holds when the
analog beamforming vectors are selected from \textit{orthogonal} codebooks
(intuitively such codebooks act as complete dictionaries in terms
of compressed sensing techniques). However, the performance of the
analog beam selection technique based on the received power can be
further improved because the corresponding effective channel $\mathbf{H}_{E}$
is not necessarily well-conditioned \cite{Golub1996,Tse2005}. In
other words, the factor dominating the performance of hybrid beamforming
is the singular values of $\mathbf{H}_{E}$ rather than the received
power. 

To find an effective channel yielding the maximum throughput, one
can reserve a few more candidates for the analog beamforming vectors
corresponding to the large received power levels. Then, the subset
of these candidates yielding the maximum throughput will provide the
optimal solution to the hybrid beamforming problem. Again it is evident
that the computational complexity exponentially increases as the size
of the enlarged candidate set. Consequently, we have a strong motivation
to find a relationship between the observations for the analog beam
selection and key parameters of the hybrid beamforming gain. The
relationship can be used to facilitate the process of determining
the optimal analog beamforming vectors. First let us ask, what is
actually the key quantity or parameter that leads to hybrid beamforming
gain? Depending on the SNR, we find that it is either the Frobenius
norm of the effective channel $\mathbf{H}_{E}$ or the absolute value
of the determinant of $\mathbf{H}_{E}$. $\mathbf{H}_{E}$ can be
regarded as a coupling of the channel and analog beamforming on both
sides. Such coupling coefficients can be obtained by transmitting
known pilot signals and used for not only the analog beam selection
but also constructing alternatives for $\mathbf{H}_{E}$. Accordingly,
estimates of the coupling coefficients yielding the maximum value
of the key parameters give us the necessary information to optimally
select the analog beamforming vectors.

The \textbf{problem statements} and \textbf{contributions} of the
proposed algorithm are summarized as follows: 
\begin{enumerate}
\item Most hybrid beamforming methods in the literature are implemented
based on \textit{explicit} CSI ($\mathbf{H}$) but ignore the complexity
of channel estimation or SVD. Therefore, this paper presents a method
that uses \textit{implicit }CSI (the received correlated pilots) to
implement the hybrid beamforming and omit channel estimation for large
antenna arrays. 
\item To simplify the joint problem of the precoder and combiner, some previously
proposed methods decouple these two by the assumption that either
the transmitter or receiver employs fully digital beamforming. However,
the assumption is not necessary. This paper shows that the precoder
and combiner can be implemented simultaneously with reasonable complexity
based on the estimates of the received power levels. 
\item Compared with existing approaches, we formulate a different optimization
problem by using the key parameters of hybrid beamforming gain, which
significantly alleviates the complexity of hybrid beamforming problem.
\end{enumerate}
The rest of the paper is organized as follows: Section II describes
the system and mmWave frequency-selective fading channel models. Section
III states the objective of hybrid beamforming problem. Based on the
objective function, a hybrid beamforming algorithm based on implicit
CSI is presented in Section IV. A theoretical analysis in terms of
statistical properties of effective noise occurring in the proposed
method is detailed in Section V. To support this analysis, simulation
results are presented in Section VI, and we conclude our work in Section
VII.

We use the following notations throughout this paper.

\begin{tabular}{cl}
$a$ & A scalar\tabularnewline
$\mathbf{a}$ & A column vector\tabularnewline
$\mathbf{A}$ & A matrix\tabularnewline
$\mathcal{A}$ & A set\tabularnewline
$\left[\mathbf{A}\right]_{n,n}$ & The $n^{\text{th}}$ diagonal element of $\mathbf{A}$\tabularnewline
$\left[\mathbf{A}\right]_{:,1:N}$ & The first $N$ column vectors of $\mathbf{A}$\tabularnewline
$\left[\mathbf{A}\right]_{1:N,1:N}$ & The $N\times N$ submatrix extracted from the \tabularnewline
 & upper-left corner of $\mathbf{A}$\tabularnewline
$\mathbf{A}^{*}$ & The complex conjugate of $\mathbf{A}$\tabularnewline
$\mathbf{A}^{H}$ & The Hermitian transpose of $\mathbf{A}$\tabularnewline
$\mathbf{A}^{T}$ & The transpose of $\mathbf{A}$\tabularnewline
$\left\Vert \mathbf{A}\right\Vert _{F}$ & The Frobenius norm of $\mathbf{A}$\tabularnewline
$\text{det}(\mathbf{A})$ & The determinant of $\mathbf{A}$\tabularnewline
$\text{vec}(\mathbf{A})$ & The vectorization of $\mathbf{A}$\tabularnewline
$[\mathbf{A}\,|\,\mathbf{B}]$ & The horizontal concatenation\tabularnewline
$\mathbf{A}\otimes\mathbf{B}$ & The Kronecker product of $\mathbf{A}$ and $\mathbf{B}$\tabularnewline
$\mathfrak{R}(\mathbf{A})$, $\mathfrak{I}(\mathbf{A})$ & The real (or imaginary) part of $\mathbf{A}$\tabularnewline
$\mathbf{I}_{N}$ & The $N\times N$ identity matrix\tabularnewline
$\mathbf{0}_{N\times M}$ & The $N\times M$ zero matrix\tabularnewline
$\text{E}\left[\cdot\right]$ & The expectation operator\tabularnewline
\end{tabular}

\section{System Model\label{sec:System_model}}

\begin{figure*}[t]
\centering{}\includegraphics[scale=0.85]{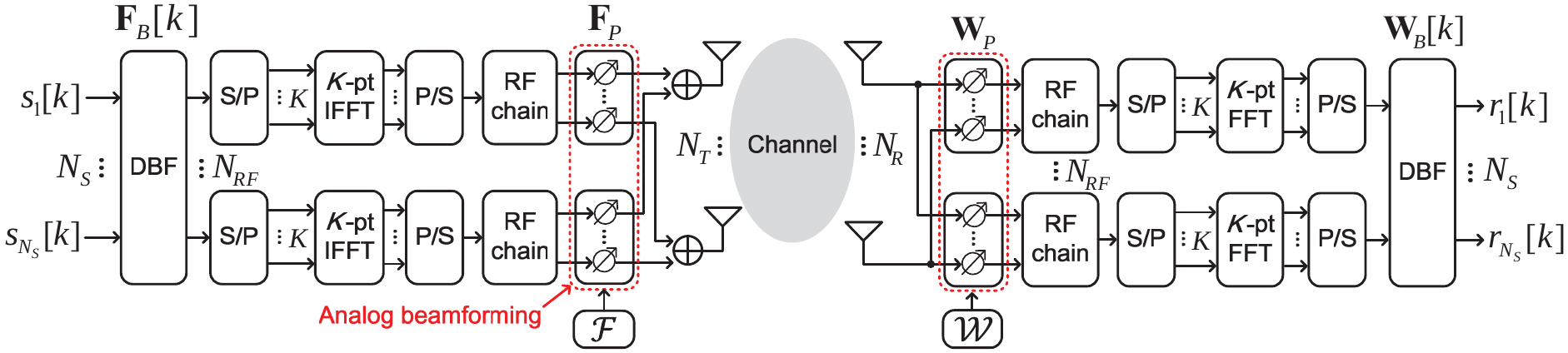}\caption{A MIMO-OFDM transceiver has hybrid analog and digital beamforming
(DBF) structures on both sides, where each analog beamforming vector
is represented by multiple phase shifters connecting to one RF chain.
\label{fig:System-diagram.} }
\end{figure*}

A system has a transmitter with a uniform linear array (ULA) of $N_{T}$
elements and wants to communicate $N_{S}$ OFDM data streams to a
receiver with an $N_{R}$-element ULA as shown in Fig. \ref{fig:System-diagram.}.
At the transmitter, the $N_{T}$ antenna elements connect to a precoder
$\mathbf{F}_{P}\mathbf{F}_{B}[k]$ at subcarrier $k=1,\cdots,K$,
where $\mathbf{F}_{P}\in\mathbb{C}^{N_{T}\times N_{RF}}$ is the analog
beamforming matrix implemented in passband as part of the RF front
end and $\mathbf{F}_{B}[k]\in\mathbb{C}^{N_{RF}\times N_{S}}$ is
the digital beamforming matrix in baseband. The value $K$ specifies
the number of subcarriers in one OFDM symbol, and $N_{RF}$ denotes
the number of available RF chains at both the transmitter and receiver.
High implementation costs and power consumption impose hardware constraints
on the analog beamforming ($\mathbf{F}_{P}$) so that it has fewer
degrees of freedom than the digital beamforming ($\mathbf{F}_{B}[k]$).
Specifically, first, $\mathbf{F}_{P}$ should be a constant matrix
within (at least) one OFDM symbol, which requires certain coherence
time of the channel. Second, the entries of $\mathbf{F}_{P}$ have
equal magnitude because analog beamformers are typically implemented
by delay elements in the RF front end. The $N_{RF}$ analog beamforming
vectors of $\mathbf{F}_{P}$ are selected from a predefined codebook
$\mathcal{F}=\{\tilde{\mathbf{f}}_{n_{f}}\in\mathbb{C}^{N_{T}\times1},n_{f}=1,\cdots,N_{F}\}$
with the $n_{f}^{\text{th}}$ member given by \cite{Liberti1999}
\begin{multline}
\tilde{\mathbf{f}}_{n_{f}}=\frac{1}{\sqrt{N_{T}}}\left[1,e^{j\frac{2\pi}{\lambda_{0}}\text{sin}(\phi_{T,n_{f}})\Delta_{d}},\cdots,\right.\\
\left.e^{j\frac{2\pi}{\lambda_{0}}\text{sin}(\phi_{T,n_{f}})(N_{T}-1)\Delta_{d}}\right]^{T},\label{eq: f}
\end{multline}
where $\phi_{T,n_{f}}$ stands for the $n_{f}^{\text{th}}$ candidate
for the steering angles at the transmitter, $\Delta_{d}=\lambda_{0}/2$
is the distance between two neighboring antennas, and $\lambda_{0}$
is the wavelength at the carrier frequency. At the receiver, the combiner
$\mathbf{W}_{P}\mathbf{W}_{B}[k]$ has a similar structure as the
precoder, where $\mathbf{W}_{P}\in\mathbb{C}^{N_{R}\times N_{RF}}$
and $\mathbf{W}_{B}[k]\in\mathbb{C}^{N_{RF}\times N_{S}}$ are the
analog and digital beamforming matrices respectively. Also, the columns
of $\mathbf{W}_{P}$ are selected from the other codebook $\mathcal{W}=\{\tilde{\mathbf{w}}_{n_{w}}\in\mathbb{C}^{N_{R}\times1},n_{w}=1,\cdots,N_{W}\}$,
where the members of $\mathcal{W}$ can be generated by the same rule
as (\ref{eq: f}).

Via a coupling of the precoder, combiner, and a frequency-selective
fading channel $\mathbf{H}[k]\in\mathbb{C}^{N_{R}\times N_{T}}$,
the received signal $\mathbf{r}[k]\in\mathbb{C}^{N_{S}\times1}$ at
subcarrier $k$ can be written as 
\begin{equation}
\begin{alignedat}{1}\mathbf{r}[k] & =\mathbf{W}_{B}^{H}[k]\mathbf{W}_{P}^{H}\mathbf{H}[k]\underset{\underline{\mathbf{s}}[k]}{\underbrace{\mathbf{F}_{P}\mathbf{F}_{B}[k]\mathbf{s}[k]}}+\underset{\underline{\mathbf{n}}[k]}{\underbrace{\mathbf{W}_{B}^{H}[k]\mathbf{W}_{P}^{H}\mathbf{n}[k]}}\\
 & =\mathbf{W}_{B}^{H}[k]\mathbf{W}_{P}^{H}\mathbf{H}[k]\underline{\mathbf{s}}[k]+\underline{\mathbf{n}}[k],
\end{alignedat}
\label{eq: r}
\end{equation}
where $\mathbf{s}[k]\in\mathbb{C}^{N_{S}\times1}$ is the transmitted
signal vector whose covariance matrix is $\mathbf{R}_{s}=\text{E}[\mathbf{s}[k]\mathbf{s}^{H}[k]]$,
and $\mathbf{n}[k]\in\mathbb{C}^{N_{R}\times1}$ is an $N_{R}$-dimensional
circularly symmetric complex Gaussian (CSCG) random vector with mean
$\boldsymbol{0}_{N_{R}\times1}$ and covariance matrix $\sigma_{n}^{2}\mathbf{I}_{N_{R}}$,
$\mathbf{n}[k]\sim\mathcal{CN}(\boldsymbol{0}_{N_{R}\times1},\sigma_{n}^{2}\mathbf{I}_{N_{R}})$.
Furthermore, the precoded transmitted signal vector $\underline{\mathbf{s}}[k]\in\mathbb{C}^{N_{T}\times1}$
and combined noise vector $\underline{\mathbf{n}}[k]\in\mathbb{C}^{N_{S}\times1}$
are enforced to satisfy the following two conditions respectively:
(1) constant transmit power on each subcarrier\footnote{We consider a stricter condition that the constant power is allocated
per subcarrier instead of per OFDM symbol for the sake of low complexity.}, and (2) the entries of $\underline{\mathbf{n}}[k]$ remain i.i.d.,
i.e.,
\begin{flalign}
\text{tr}(\mathbf{R}_{\underline{s}}) & =\text{tr}\left(\mathbf{F}_{P}\mathbf{F}_{B}[k]\mathbf{R}_{s}\mathbf{F}_{B}^{H}[k]\mathbf{F}_{P}^{H}\right)=\text{tr}(\mathbf{R}_{s}),\label{eq: constraint_1}\\
\mathbf{R}_{\underline{n}} & =\sigma_{n}^{2}\mathbf{W}_{B}^{H}[k]\mathbf{W}_{P}^{H}\mathbf{W}_{P}\mathbf{W}_{B}[k]=\sigma_{n}^{2}\mathbf{I}_{N_{S}},\label{eq: constraint_2}
\end{flalign}
where $\mathbf{R}_{\underline{s}}=\text{E}[\underline{\mathbf{s}}[k]\underline{\mathbf{s}}^{H}[k]]$
and $\mathbf{R}_{\underline{n}}=\text{E}[\underline{\mathbf{n}}[k]\underline{\mathbf{n}}^{H}[k]]$
are the covariance matrices of $\underline{\mathbf{s}}[k]$ and $\underline{\mathbf{n}}[k]$
respectively. These two equations, (\ref{eq: constraint_1}) and (\ref{eq: constraint_2}),
can also be regarded as the power constraints on the precoder and
combiner.

\begin{figure}[t]
\begin{centering}
\includegraphics[scale=0.51]{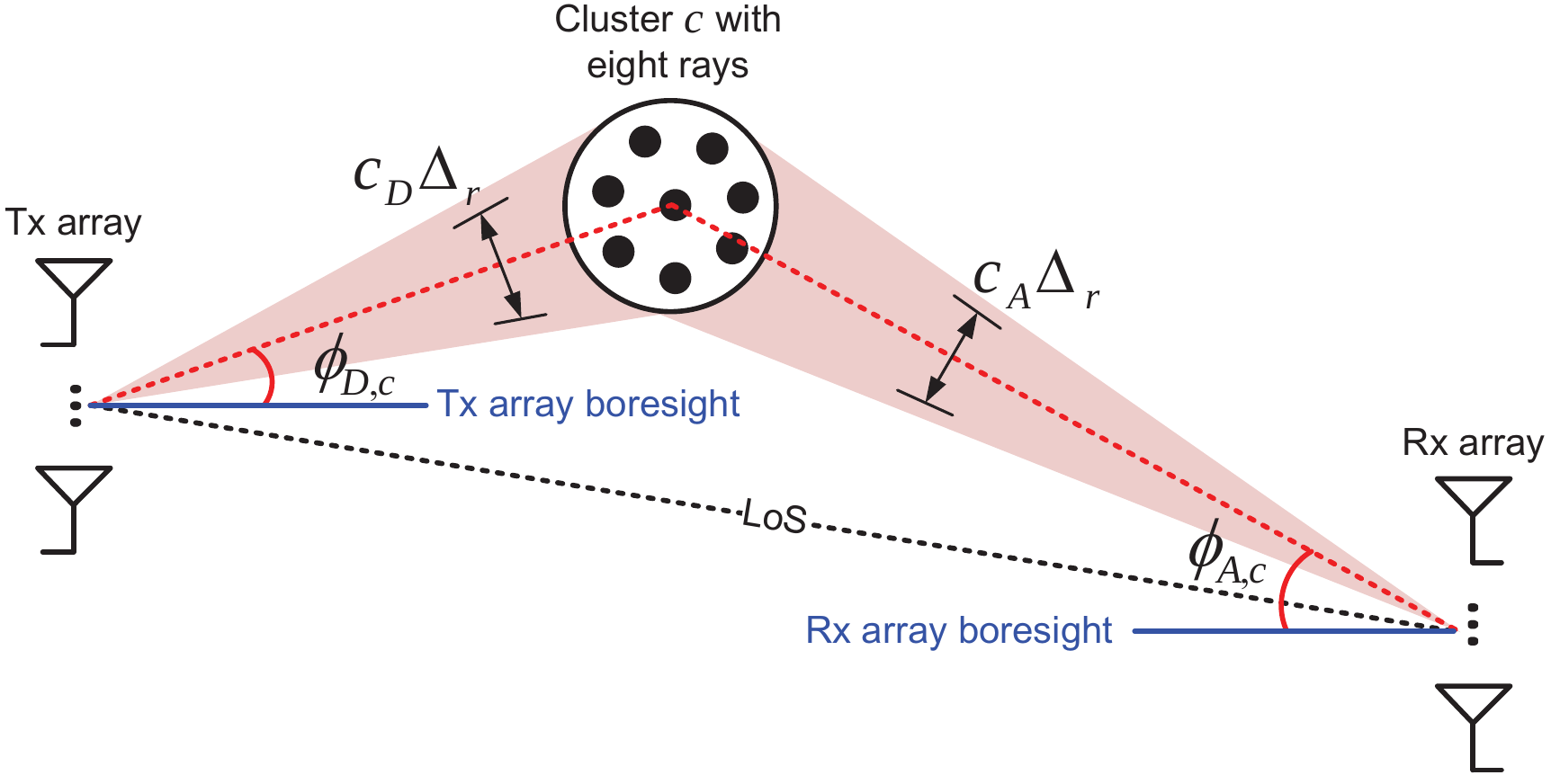} 
\par\end{centering}
\caption{An example of the angular spread in a cluster characterized by its
mean values ($\phi_{D,c}$ and $\phi_{A,c}$) and intra-cluster angular
spreads ($c_{D}\Delta_{r}$ and $c_{A}\Delta_{r}$). \label{fig:An-example-of}}
\end{figure}

The properties of mmWave channels have been widely studied recently,
and simulation models have been developed accordingly \cite{Rappaport2015,3GPP38900}.
The most comprehensive one can be found in \cite{3GPP38900}. Based
on the references, a simplified cluster-based frequency-selective
fading channel has $C$ clusters and $R$ rays of each cluster\textcolor{red}{},
where $CR\geq N_{RF}$. At subcarrier $k$, the channel matrix can
be written as 
\begin{equation}
\begin{alignedat}{1}\mathbf{H}[k] & =\sqrt{\rho}\sum_{c=1}^{C}\sum_{r=1}^{R}\alpha_{c,r}\cdot e^{-\frac{j2\pi kl_{c,r}}{K}}\cdot\mathbf{a}_{A}(\phi_{A,c,r})\mathbf{a}_{D}(\phi_{D,c,r})^{H}\end{alignedat}
,
\end{equation}
where the channel characteristics are given by the following parameters:
\begin{itemize}
\item $\rho$ stands for the average received power including the transmit
power, transmit antenna gain, receive antenna gain, and path loss.
\item $\alpha_{c,r}\in\mathbb{C}$ describes the inter- and intra-cluster
path gain. The difference in power between light-of-sight (LoS) and
NLoS clusters is about $20$ dB and $\sum_{c=1}^{C}\sum_{r=1}^{R}|\alpha_{c,r}|^{2}=1$.
\item Frequency-selective properties of the channel are specified in terms
of normalized-quantized delays (i.e., delay indices measured in units
of the sampling interval) $l_{c,r}=\left\lfloor \tau_{c,r}F_{S}\right\rfloor \in\mathbb{N}_{0}$,
where $\tau_{c,r}$ and $F_{S}$ stand for the path delay and sampling
rate respectively.
\item $\phi_{D,c,r}$ is the angle of departure (AoD) of ray $r$ in cluster
$c$, see Fig. \ref{fig:An-example-of}. It is characterized by the
mean $\phi_{D,c}$, root mean square angular spread $c_{D}$, and
offset angle $\Delta_{r}$ for ray $r$, i.e.,
\begin{equation}
\begin{alignedat}{1}\phi_{D,c,r} & =\phi_{D,c}+c_{D}\Delta_{r},\end{alignedat}
\end{equation}
where $\phi_{D,c}\sim\mathcal{U}(-\tfrac{\pi}{2},\tfrac{\pi}{2})$,
$c_{D}$ and $\Delta_{r}$ are respectively given in \cite[Table 7.5-3]{3GPP38900}
and \cite[Table 7.5-6]{3GPP38900}. In the same way, one can generate
the angle of arrival (AoA) $\phi_{A,c,r}$.
\item The array response vector of the ULA to an incident plane wave at
the transmitter, $\mathbf{a}_{D}(\phi_{D,c,r})$, has $N_{T}$ entries
of equal magnitude and is a function of $\phi_{D,c,r}$ only. It can
be written as
\begin{multline}
\mathbf{a}_{D}(\phi_{D,c,r})=\frac{1}{\sqrt{N_{T}}}\left[1,e^{j\tfrac{2\pi}{\lambda_{0}}\text{sin}(\phi_{D,c,r})\Delta_{d}},\cdots,\right.\\
\left.e^{j\tfrac{2\pi}{\lambda_{0}}\text{sin}(\phi_{D,c,r})(N_{T}-1)\Delta_{d}}\right]^{T}.\label{eq: a_d}
\end{multline}
Given an AoA, the array response vector at the receiver, $\mathbf{a}_{A}(\phi_{A,c,r})$,
has a similar form as (\ref{eq: a_d}).
\end{itemize}

\section{Problem Statement \label{sec:Problem-Statement}}

In the beamforming system, the objective of the precoder $\mathbf{F}_{P}\mathbf{F}_{B}[k]\,\forall k$
and the associated combiner $\mathbf{W}_{P}\mathbf{W}_{B}[k]\,\forall k$
is to maximize the mutual information of the system subject to the
power constraints on $\mathbf{F}_{P}$, $\mathbf{W}_{P}$, $\mathbf{F}_{B}[k]$,
and $\mathbf{W}_{B}[k]$ $\forall k$. That is, we seek matrices that
solve 
\begin{equation}
\begin{gathered}\begin{gathered}{\displaystyle \underset{\mathbf{F}_{P},\mathbf{W}_{P},(\mathbf{F}_{B}[k],\mathbf{W}_{B}[k])\,\forall k}{\max}}\,\sum_{k=0}^{K-1}I(\mathbf{F}_{P},\mathbf{W}_{P},\mathbf{F}_{B}[k],\mathbf{W}_{B}[k]),\end{gathered}
\\
\text{s.t. }\begin{cases}
\mathbf{f}_{P,n_{rf}}\in\mathcal{F},\mathbf{w}_{P,n_{rf}}\in\mathcal{W}\:\:\forall n_{rf},\\
\text{tr}\left(\mathbf{F}_{P}\mathbf{F}_{B}[k]\mathbf{R}_{s}\mathbf{F}_{B}^{H}[k]\mathbf{F}_{P}^{H}\right)=\text{tr}(\mathbf{R}_{s})\,\forall k,\\
\mathbf{W}_{B}^{H}[k]\mathbf{W}_{P}^{H}\mathbf{W}_{P}\mathbf{W}_{B}[k]=\mathbf{I}_{N_{S}}\,\forall k,
\end{cases}
\end{gathered}
\label{eq: opt}
\end{equation}
where $\mathbf{f}_{P,n_{rf}}$ and $\mathbf{w}_{P,n_{rf}}$ are respectively
the $n_{rf}^{\text{th}}$ column vectors of $\mathbf{F}_{P}$ and
$\mathbf{W}_{P}$, and the last two constraints are the consequences
of (\ref{eq: constraint_1}) and (\ref{eq: constraint_2}). Assume
that $\mathbf{s}[k]$ is a CSCG random vector, i.e., $\mathbf{s}[k]\sim\mathcal{CN}(\boldsymbol{0}_{N_{S}\times1},\mathbf{R}_{s})$,
the mutual information of the system of the $k^{\text{th}}$ OFDM
subchannel is given by \cite{Ayach2014,Alkhateeb2016b,Goldsmith2003,Bolcskei2002}
\begin{multline}
I(\mathbf{F}_{P},\mathbf{W}_{P},\mathbf{F}_{B}[k],\mathbf{W}_{B}[k])\\
=\log_{2}\det\left(\mathbf{I}_{N_{S}}+\mathbf{R}_{\underline{n}}^{-1}\left(\mathbf{W}_{B}^{H}[k]\mathbf{W}_{P}^{H}\mathbf{H}[k]\mathbf{F}_{P}\mathbf{F}_{B}[k]\right)\right.\\
\left.\cdot\,\mathbf{R}_{s}\left(\mathbf{W}_{B}^{H}[k]\mathbf{W}_{P}^{H}\mathbf{H}[k]\mathbf{F}_{P}\mathbf{F}_{B}[k]\right)^{H}\right).\label{eq: I}
\end{multline}
Moreover, we denote the solution of (\ref{eq: opt}) by $(\mathbf{F}_{P,Opt},\mathbf{W}_{P,Opt},(\mathbf{F}_{B,Opt}[k],\mathbf{W}_{B,Opt}[k])\,\forall k)$. 

If explicit CSI is available, the problem of the precoder and combiner
can be solved by exploiting the SVD of the channel matrix \cite{Ayach2012,Alkhateeb2016b,Sohrabi2016}.
In the paper, we consider a more pragmatic approach that channel knowledge
is neither given nor estimated. To efficiently get the solution of
(\ref{eq: opt}) without the channel knowledge, we try an alternative
expression of (\ref{eq: opt}): given two sets $\mathcal{I_{F}}$
and $\mathcal{I_{W}}$ containing the candidates for $\mathbf{F}_{P}$
and $\mathbf{W}_{P}$, the maximum data rate of (\ref{eq: opt}) is
greater than or equal to

{\small{}\vspace{-0.3cm}
\begin{equation}
\begin{gathered}{\displaystyle \underset{\scriptsize\begin{array}{c}
\mathbf{F}_{P}\in\mathcal{I_{F}}\\
\mathbf{W}_{P}\in\mathcal{I_{W}}
\end{array}}{\max}}\underset{I_{LM}(\mathbf{F}_{P},\mathbf{W}_{P}):\text{ local maximum throughput}}{\underbrace{\left\{ \begin{gathered}\max_{(\mathbf{F}_{B}[k],\mathbf{W}_{B}[k])\,\forall k}\,\sum_{k=0}^{K-1}I(\mathbf{F}_{P},\mathbf{W}_{P},\mathbf{F}_{B}[k],\mathbf{W}_{B}[k])\\
\text{s.t. }\begin{cases}
\text{tr}\left(\mathbf{F}_{P}\mathbf{F}_{B}[k]\mathbf{R}_{s}\mathbf{F}_{B}^{H}[k]\mathbf{F}_{P}^{H}\right)=\text{tr}(\mathbf{R}_{s})\,\forall k\\
\mathbf{W}_{B}^{H}[k]\mathbf{W}_{P}^{H}\mathbf{W}_{P}\mathbf{W}_{B}[k]=\mathbf{I}_{N_{S}}\,\forall k
\end{cases}
\end{gathered}
\right\} }}.\end{gathered}
\label{eq: new}
\end{equation}
}These two versions of the hybrid beamforming problem will have the
same maximum throughput if $\mathcal{I_{F}}$ and $\mathcal{I_{W}}$
include $\mathbf{F}_{P,Opt}$ and $\mathbf{W}_{P,Opt}$ respectively. 

The reformulated problem in (\ref{eq: new}) becomes simpler because,
given $\mathbf{F}_{P}$ and $\mathbf{W}_{P}$, the \textit{inner}
problem (to obtain the local maximum throughput $I_{LM}(\mathbf{F}_{P},\mathbf{W}_{P})$)
is similar to conventional fully digital beamforming designs subject
to different power constraints \cite{VanTrees2004,Telatar1999}. In
other words, the critical issue of the hybrid beamforming is to solve
the \textit{outer} problem by an additional maximization over all
members of $\mathcal{I_{F}}$ and $\mathcal{I_{W}}$. Therefore, the
motivation is to find $\mathcal{I_{F}}$ and $\mathcal{I_{W}}$, which
ideally include $\mathbf{F}_{P,Opt}$, $\mathbf{W}_{P,Opt}$, and
perhaps few other candidates, and then select a pair $(\mathbf{F}_{P},\mathbf{W}_{P})$
from $\mathcal{I_{F}}$ and $\mathcal{I_{W}}$ that leads to the maximum
throughput.

\section{A Hybrid Beamforming Algorithm Based on Implicit CSI\label{sec:HBF_algorithm}}

In this section, we present how to use implicit CSI to find the optimal
solution to the hybrid beamforming problem. In addition, key parameters
of the hybrid beamforming gain are introduced to reduce the complexity
of the problem. 

\subsection{Initial analog beam selection\label{subsec:Initial-analog-beam}}

\begin{figure}[t]
\centering{}\includegraphics[scale=0.4]{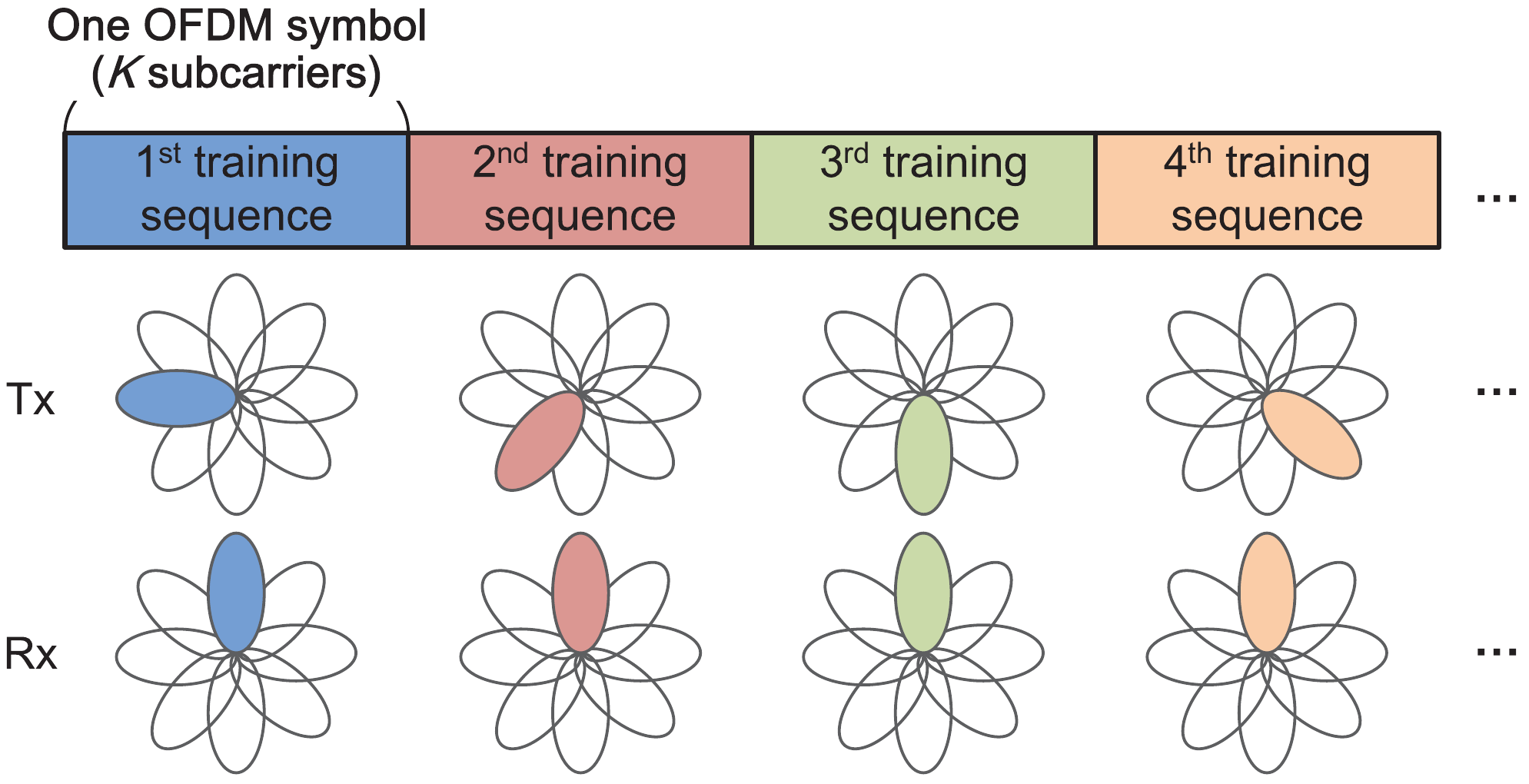}\caption{A training sequence of length $K$ is used to train a beam pair.\label{fig: CB_training}}
\end{figure}
\begin{figure*}[t]
\centering{}%
\begin{tabular}{ccc}
\subfloat[The achievable data rate by the received power of the coupling coefficients
is 2.5 bit/s/Hz.]{\includegraphics[scale=0.3]{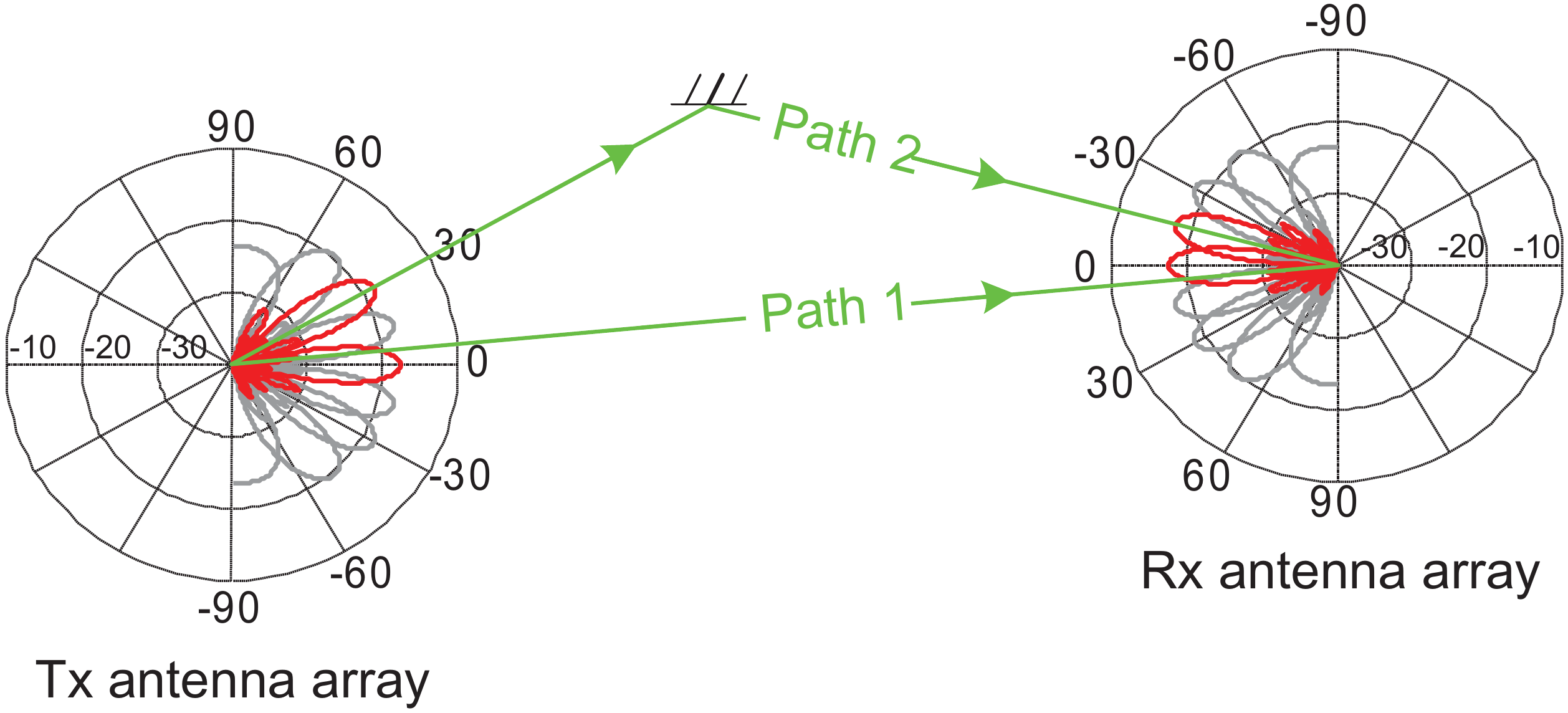}

} & $\qquad$  & \subfloat[The achievable data rate by a linear combination of two analog beamforming
vectors is 3 bit/s/Hz.]{\includegraphics[scale=0.3]{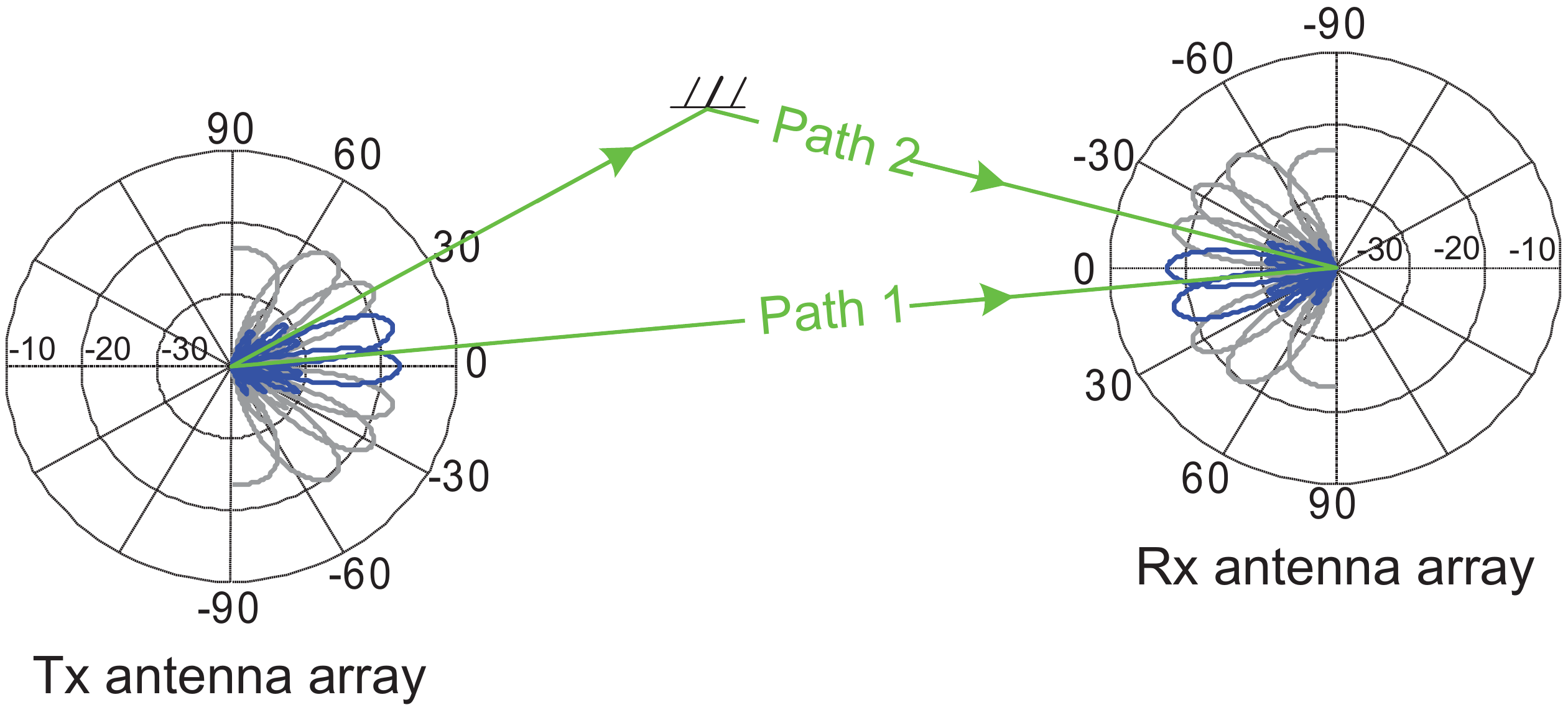}

}\tabularnewline
\end{tabular}\caption{A typical example of analog beam selection by two different approaches.
In the simplified two-path channel model, the AoDs are $\{5^{\circ},30^{\circ}\}$,
the AoAs are $\{5^{\circ},-15^{\circ}\}$, and the difference in path
attenuation between path one and two amounts to 10 dB. \label{fig:example}}
\end{figure*}
To begin with, let us see how to obtain the sets $\mathcal{I_{F}}$
and $\mathcal{I_{W}}$ in (\ref{eq: new}) from the given codebooks
$\mathcal{F}$ and $\mathcal{W}$. We call this step \textit{initial
analog beam selection}. By transmitting known pilot signals, we have
the received pilot signals used for the initial analog beam selection.
However, these received signals include the effect of analog beamforming
because the hardware-constrained analog beamforming matrices $\mathbf{F}_{P}$
and $\mathbf{W}_{P}$ cannot be replaced by the identity matrices.
As a result, one can simply assume that all the members of the codebooks
$\mathcal{F}=\{\tilde{\mathbf{f}}_{n_{f}},n_{f}=1,\cdots,N_{F}\}$
and $\mathcal{W}=\{\tilde{\mathbf{w}}_{n_{w}},n_{w}=1,\cdots,N_{W}\}$
are trained by transmitting a training sequence $\{s[k]\}_{k=0}^{K-1}$
that satisfies $|s[k]|^{2}=1\,\forall k$, as shown in Fig. \ref{fig: CB_training}.
Then an observation used for the analog beam selection at subcarrier
$k$ for a specific beam pair $(\tilde{\mathbf{f}}_{n_{f}},\tilde{\mathbf{w}}_{n_{w}})$
can be acquired by correlating the $k^{\text{th}}$ received pilot
with its transmitted signal
\begin{equation}
\begin{aligned}y_{n_{w},n_{f}}[k] & =\frac{s^{*}[k]}{|s[k]|^{2}}\underset{\text{received pilot signal}}{\underbrace{\left(\tilde{\mathbf{w}}_{n_{w}}^{H}\mathbf{H}[k]\tilde{\mathbf{f}}_{n_{f}}s[k]+\tilde{\mathbf{w}}_{n_{w}}^{H}\mathbf{n}[k]\right)}}\\
 & =\tilde{\mathbf{w}}_{n_{w}}^{H}\mathbf{H}[k]\tilde{\mathbf{f}}_{n_{f}}+\underset{z_{n_{w},n_{f}}[k]}{\underbrace{\frac{s^{*}[k]}{|s[k]|^{2}}\tilde{\mathbf{w}}_{n_{w}}^{H}\mathbf{n}[k]}}\\
 & =\tilde{\mathbf{w}}_{n_{w}}^{H}\mathbf{H}[k]\tilde{\mathbf{f}}_{n_{f}}+z_{n_{w},n_{f}}[k].
\end{aligned}
\label{eq: y}
\end{equation}
Similar observations become available on all subcarriers and the effective
noise $z_{n_{w},n_{f}}[k]\sim\mathcal{CN}(0,\sigma_{n}^{2})$ still
has a Gaussian distribution with mean zero and variance $\sigma_{n}^{2}$.
\textcolor{blue}{{} }Also, $z_{n_{w},n_{f}}[k]$ is expressed as a function
of $n_{w}$ and $n_{f}$ as the noise vector $\mathbf{n}[k]$ is random
for a trained analog beam pair $(\tilde{\mathbf{f}}_{n_{f}},\tilde{\mathbf{w}}_{n_{w}})$.
The observation $y_{n_{w},n_{f}}[k]$ can be viewed as implicit CSI,
which is a coupling coefficient corresponding to a pair of analog
beamforming vectors selected on both sides of the channel.

Borrowing the idea from our previous works in \cite{Chiang2017_ICASSP,Chiang2017_WSA},
it shows that when $\mathcal{F}$ and $\mathcal{W}$ are orthogonal
codebooks\footnote{To be formal, an orthogonal codebook $\mathcal{F}$ satisfies $\frac{\left\langle \tilde{\mathbf{f}}_{i},\tilde{\mathbf{f}}_{j}\right\rangle }{\left\Vert \tilde{\mathbf{f}}_{i}\right\Vert _{2}\left\Vert \tilde{\mathbf{f}}_{j}\right\Vert _{2}}=\begin{cases}
0, & i\neq j\\
1, & i=j
\end{cases},$ where $\left\langle \tilde{\mathbf{f}}_{i},\tilde{\mathbf{f}}_{j}\right\rangle $
denotes the inner product of the two vectors.}, the sum of the power of $K$ observations in one OFDM symbol can
be directly used for the analog beam selection. Consequently, $M$
analog beam pairs (assume that $M\geq N_{RF}$, which will be explained
later) can be selected individually and sequentially according to
the sorted received energy estimates 

\begin{equation}
\begin{gathered}(\hat{\mathbf{f}}_{m},\hat{\mathbf{w}}_{m})={\displaystyle \underset{\scriptsize\begin{array}{c}
\tilde{\mathbf{f}}_{n_{f}}\in\mathcal{F}\backslash\mathcal{F}',\tilde{\mathbf{w}}_{n_{w}}\in\mathcal{W}\backslash\mathcal{W}'\end{array}}{\arg\,\max}}\sum_{k=0}^{K-1}\left|y_{n_{w},n_{f}}[k]\right|^{2},\end{gathered}
\label{eq: ABF}
\end{equation}
where $m=1,\cdots,M$, $\mathcal{F}'=\{\hat{\mathbf{f}}_{n},n=1,\cdots,m-1\}$
and $\mathcal{W}'=\{\hat{\mathbf{w}}_{n},n=1,\cdots,m-1\}$ are the
sets consisting of the selected analog beamforming vectors from iteration
$1$ to $m-1$. 

We assume that $M\geq N_{RF}$ for the reason that the first $N_{RF}$
selected analog beam pairs according to the sorted values of $\sum_{k=0}^{K-1}|y_{n_{w},n_{f}}[k]|^{2}$,
where $n_{w}=1,\cdots,N_{W}$ and $n_{f}=1,\cdots,N_{F}$, may not
be equal to the optimal solution ($\mathbf{F}_{P,Opt}$ and $\mathbf{W}_{P,Opt}$)
because we do not yet consider the effect of digital beamforming during
the analog beam selection phase. In addition, (\ref{eq: ABF}) is
derived from the assumption that $\mathcal{F}$ and $\mathcal{W}$
are orthogonal codebooks. To find the optimal solution ($\mathbf{F}_{P,Opt}$
and $\mathbf{W}_{P,Opt}$) for any type of codebook (orthogonal or
non-orthogonal), one has to further take into account linear combinations
of $N_{RF}$ analog beamforming vectors selected from $\{\hat{\mathbf{f}}_{m}\,\forall m\}$
and $\{\hat{\mathbf{w}}_{m}\,\forall m\}$ with coefficients in digital
beamforming. To this end, we define two sets $\mathcal{I_{F}}$
and $\mathcal{I_{W}}$ consisting of all combinations of $N_{RF}$
members chosen from $\{\hat{\mathbf{f}}_{m},m=1,\cdots,M\}$ and $\{\hat{\mathbf{w}}_{m},m=1,\cdots,M\}$,
respectively, which can be written as
\begin{equation}
\begin{alignedat}{1}\mathcal{I_{F}} & =\{\overline{\mathbf{F}}_{P,i_{f}},i_{f}=1,\cdots,I_{F}\},\\
\mathcal{I_{W}} & =\{\overline{\mathbf{W}}_{P,i_{w}},i_{w}=1,\cdots,I_{W}\},
\end{alignedat}
\label{eq: beam space}
\end{equation}
where the cardinality $I_{F}=I_{W}=\binom{M}{N_{RF}}$ of both sets
is given by the binomial coefficient. The notations $\overline{\mathbf{F}}_{P,i_{f}}$
and $\overline{\mathbf{W}}_{P,i_{w}}$ respectively denote the $i_{f}^{\text{th}}$
and $i_{w}^{\text{th}}$ candidates for the analog beamforming matrices
$\mathbf{F}_{P}$ and $\mathbf{W}_{P}$. When $M$ becomes large,
there is a high probability that $\mathcal{I_{F}}$ and $\mathcal{I_{W}}$
include the global optimum solution ($\mathbf{F}_{P,Opt}$ and $\mathbf{W}_{P,Opt}$).\\

\textit{Schematic example}: To illustrate the concept, let us consider
a scenario with $N_{T}=N_{R}=8$ antenna elements, codebook sizes
$N_{F}=N_{W}=8$, the same orthogonal codebook $\mathcal{F}=\mathcal{W}$
at the transmitter and receiver with the candidates for steering angles
given by $\{-90^{\circ}\:(\text{or }90^{\circ}),-48.59^{\circ},-30^{\circ},-14.48^{\circ},0^{\circ},$\\
 $14.48^{\circ},30^{\circ},48.59^{\circ}\}$, and $N_{RF}=2$ available
RF chains to transmit $N_{S}=2$ data streams at $\text{SNR}=5\text{ dB}$.

\begin{figure}[t]
\centering{}\hspace*{-0.1cm}\includegraphics[scale=0.33]{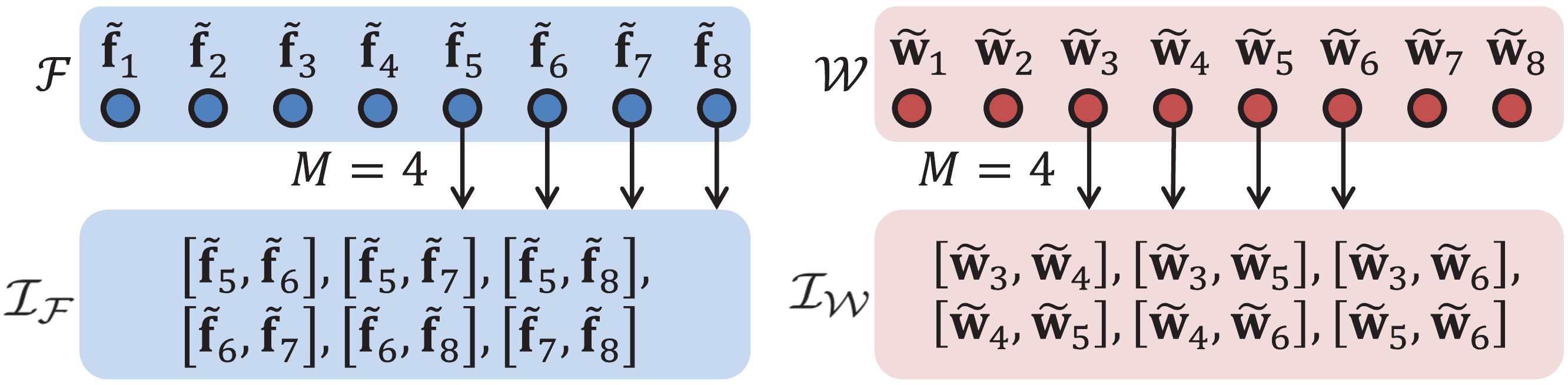}\caption{An example of the codebooks, $\mathcal{F}$ and $\mathcal{W}$, and
the sets $\mathcal{I_{F}}$ and $\mathcal{I_{W}}$ consisting of $\binom{M}{N_{RF}}=\binom{4}{2}=6$
candidates for $\mathbf{F}_{P}$ and $\mathbf{W}_{P}$ respectively.\label{fig:An-example-of-1} }
\end{figure}

The channel realization as depicted in Fig. \ref{fig:example} has
two paths. In Fig. \ref{fig:example}(a), two analog beam pairs selected
according to (\ref{eq: ABF}) steer towards these two paths (highlighted
in red). Before digital beamforming comes into play, the analog beamforming
vectors would be used with the same weighting. If more than $N_{RF}=2$
analog beam pairs are reserved, more options with digital beamforming
can be explored. In this example, with $M=4$, we have $I_{F}=I_{W}=\binom{M}{N_{RF}}=\binom{4}{2}=6$
members in both $\mathcal{I_{F}}$ and $\mathcal{I_{W}}$, see Fig.
\ref{fig:An-example-of-1}. We enumerate them explicitly as 
\[
\begin{alignedat}{1}\mathcal{I_{F}} & =\{\overline{\mathbf{F}}_{P,i_{f}},i_{f}=1,\cdots,6\},\\
\mathcal{I_{W}} & =\{\overline{\mathbf{W}}_{P,i_{w}},i_{w}=1,\cdots,6\}.
\end{alignedat}
\]
For instance, $\overline{\mathbf{F}}_{P,1}=[\tilde{\mathbf{f}}_{5},\tilde{\mathbf{f}}_{6}]$
and $\overline{\mathbf{W}}_{P,1}=[\tilde{\mathbf{w}}_{3},\tilde{\mathbf{w}}_{4}]$.
Therefore, one can try $36$ pairs, $\{(\overline{\mathbf{F}}_{P,i_{f}},\overline{\mathbf{W}}_{P,i_{w}})\,|\,\forall i_{f},i_{w}\}$,
to determine the optimal weights of digital beamformers and the corresponding
analog beamforming matrices, which will be detailed in the following
subsections. In general, there will be a competition between spatial
multiplexing gain over different propagation paths and power gain
available from the dominant path. In this case, the two analog beam
pairs highlighted in blue in Fig. \ref{fig:example}(b) steer to the
dominant path and lead to higher spectral efficiency. However, which
beamforming strategy yields higher throughput in any specific case
is not clear beforehand.

\subsection{Digital beamforming}

After the initial analog beam selection, we are in possession of the
two sets $\mathcal{I_{F}}$ and $\mathcal{I_{W}}$ that contain the
candidates for $\mathbf{F}_{P}$ and $\mathbf{W}_{P}$, and the objective
is to efficiently find the optimal solution. Before going into the
detail of our proposed scheme, let us review the relationship between
the analog and digital beamforming. Given one particular choice $(\overline{\mathbf{F}}_{P,i_{f}},\overline{\mathbf{W}}_{P,i_{w}})$
selected from the candidate sets $\mathcal{I_{F}}$ and $\mathcal{I_{W}}$,
it is clear that the goal of digital beamforming is to maximize the
local maximum throughput $I_{LM}(\overline{\mathbf{F}}_{P,i_{f}},\overline{\mathbf{W}}_{P,i_{w}})$,
as defined in (\ref{eq: new}), with the objective function expressed
as
\begin{equation}
\begin{gathered}\max_{(\mathbf{F}_{B}[k],\mathbf{W}_{B}[k])\,\forall k}\,\sum_{k=0}^{K-1}I(\overline{\mathbf{F}}_{P,i_{f}},\overline{\mathbf{W}}_{P,i_{w}},\mathbf{F}_{B}[k],\mathbf{W}_{B}[k])\\
=\sum_{k=0}^{K-1}\,\max_{\mathbf{F}_{B}[k],\mathbf{W}_{B}[k]}\,I(\overline{\mathbf{F}}_{P,i_{f}},\overline{\mathbf{W}}_{P,i_{w}},\mathbf{F}_{B}[k],\mathbf{W}_{B}[k]),
\end{gathered}
\end{equation}
where $I(\overline{\mathbf{F}}_{P,i_{f}},\overline{\mathbf{W}}_{P,i_{w}},\mathbf{F}_{B}[k],\mathbf{W}_{B}[k])$
is given by (\ref{eq: I}) with $\mathbf{F}_{P}=\overline{\mathbf{F}}_{P,i_{f}}$
and $\mathbf{W}_{P}=\overline{\mathbf{W}}_{P,i_{w}}$. As a result,
the digital beamforming problem at subcarrier $k$ can be formulated
as a throughput maximization problem subject to the power constraints,
which can be stated as

\vspace{-0.3cm}

{\small{}
\begin{equation}
\begin{gathered}(\overline{\mathbf{F}}_{B,i}[k],\overline{\mathbf{W}}_{B,i}[k])={\displaystyle \underset{\mathbf{F}_{B}[k],\mathbf{W}_{B}[k]}{\arg\,\max}}I(\overline{\mathbf{F}}_{P,i_{f}},\overline{\mathbf{W}}_{P,i_{w}},\mathbf{F}_{B}[k],\mathbf{W}_{B}[k])\\
\text{s.t. }\begin{cases}
\text{tr}\left(\overline{\mathbf{F}}_{P,i_{f}}\mathbf{F}_{B}[k]\mathbf{R}_{s}\mathbf{F}_{B}^{H}[k]\overline{\mathbf{F}}_{P,i_{f}}^{H}\right)=\text{tr}(\mathbf{R}_{s}),\\
\mathbf{W}_{B}^{H}[k]\overline{\mathbf{W}}_{P,i_{w}}^{H}\overline{\mathbf{W}}_{P,i_{w}}\mathbf{W}_{B}[k]=\mathbf{I}_{N_{S}},
\end{cases}
\end{gathered}
\label{eq:constrained_DBF}
\end{equation}
}where $i=(i_{f}-1)I_{W}+i_{w}$ is an index specifying the combined
members of $\mathcal{I_{F}}$ and $\mathcal{I_{W}}$. To proceed,
we take advantage of the mathematical results in Appendix \ref{sec:Derivation-of-DBF}
and have
\begin{align}
\overline{\mathbf{F}}_{B,i}[k] & =(\overline{\mathbf{F}}_{P,i_{f}}^{H}\overline{\mathbf{F}}_{P,i_{f}})^{-0.5}\left[\mathbf{V}_{E,i}[k]\right]_{:,1:N_{S}},\\
\overline{\mathbf{W}}_{B,i}[k] & =(\overline{\mathbf{W}}_{P,i_{w}}^{H}\overline{\mathbf{W}}_{P,i_{w}})^{-0.5}\left[\mathbf{U}_{E,i}[k]\right]_{:,1:N_{S}},
\end{align}
where the columns of $\mathbf{V}_{E,i}[k]$ and $\mathbf{U}_{E,i}[k]$
are respectively the right- and left-singular vectors of the effective
channel matrix defined by
\begin{equation}
\begin{alignedat}{1}\mathbf{H}_{E,i}[k] & \triangleq(\overline{\mathbf{W}}_{P,i_{w}}^{H}\overline{\mathbf{W}}_{P,i_{w}})^{-0.5}\overline{\mathbf{W}}_{P,i_{w}}^{H}\mathbf{H}[k]\overline{\mathbf{F}}_{P,i_{f}}(\overline{\mathbf{F}}_{P,i_{f}}^{H}\overline{\mathbf{F}}_{P,i_{f}})^{-0.5}.\end{alignedat}
\label{eq:H_E}
\end{equation}

\subsection{Key parameters of hybrid beamforming gain\label{subsec:Key-parameter-of}}

As mentioned above, given a pair of members selected from $\mathcal{I_{F}}$
and $\mathcal{I_{W}}$, $(\overline{\mathbf{F}}_{P,i_{f}},\overline{\mathbf{W}}_{P,i_{w}})$,
we have the corresponding optimal digital beamforming matrices $(\overline{\mathbf{F}}_{B,i}[k],\overline{\mathbf{W}}_{B,i}[k])\:\forall k$.
Accordingly, the local maximum throughput is given by
\begin{equation}
\begin{alignedat}{1} & I_{LM}(\overline{\mathbf{F}}_{P,i_{f}},\overline{\mathbf{W}}_{P,i_{w}})\\
 & =\sum_{k=0}^{K-1}I(\overline{\mathbf{F}}_{P,i_{f}},\overline{\mathbf{W}}_{P,i_{w}},\overline{\mathbf{F}}_{B,i}[k],\overline{\mathbf{W}}_{B,i}[k])\\
 & =\sum_{k=0}^{K-1}\sum_{n_{s}=1}^{N_{S}}\log_{2}\left(1+\frac{1}{\sigma_{n}^{2}}\left[\boldsymbol{\mathbf{\Sigma}}_{E,i}^{2}[k]\right]_{n_{s},n_{s}}\left[\mathbf{R}_{s}\right]_{n_{s},n_{s}}\right),
\end{alignedat}
\label{eq: I_LM}
\end{equation}
where the diagonal elements of $\boldsymbol{\mathbf{\Sigma}}_{E,i}[k]$
are the singular values of the effective channel $\mathbf{H}_{E,i}[k]\overset{\text{SVD}}{=}\mathbf{U}_{E,i}[k]\boldsymbol{\mathbf{\Sigma}}_{E,i}[k]\mathbf{V}_{E,i}^{H}[k]$.
Based on the candidate set $\{(\overline{\mathbf{F}}_{P,i_{f}},\overline{\mathbf{W}}_{P,i_{w}})\,|\,\forall i_{f},i_{w}\}$,
the pair leading to the maximum throughput provides the best approximation
of the global optimal analog beamforming matrices, that is, the solution
to the hybrid beamforming problem in (\ref{eq: new}), written as
\begin{equation}
\begin{alignedat}{1}\left(\hat{\mathbf{F}}_{P},\hat{\mathbf{W}}_{P}\right) & ={\displaystyle \underset{\scriptsize\begin{array}{c}
\overline{\mathbf{F}}_{P,i_{f}}\in\mathcal{I_{F}},\overline{\mathbf{W}}_{P,i_{w}}\in\mathcal{I_{W}}\end{array}}{\arg\,\max}}\,I_{LM}(\overline{\mathbf{F}}_{P,i_{f}},\overline{\mathbf{W}}_{P,i_{w}}).\end{alignedat}
\label{eq: max}
\end{equation}
However, this way of solving the problem requires the SVD of $\{\mathbf{H}_{E,i}[k]\}_{k=0}^{K-1}$
to obtain $I_{LM}(\overline{\mathbf{F}}_{P,i_{f}},\overline{\mathbf{W}}_{P,i_{w}})$
for each pair, which means that we have to repeat the calculation
as many as $\binom{M}{N_{RF}}^{2}$ times. 

Alternatives that can reduce the potentially large computational burden
are necessary. We ask ourselves what are the crucial parameter(s)
or indicator(s) that actually determine the throughput. To answer
this question, let $\mathbf{R}_{s}=\frac{1}{N_{S}}\mathbf{I}_{N_{S}}$
(equal power allocation) so that the maximum achievable throughput
at subcarrier $k$  becomes
\begin{equation}
\begin{alignedat}{1} & I\left(\overline{\mathbf{F}}_{P,i_{f}},\overline{\mathbf{W}}_{P,i_{w}},\overline{\mathbf{F}}_{B,i}[k],\overline{\mathbf{W}}_{B,i}[k]\right)\\
 & \qquad\qquad=\sum_{n_{s}=1}^{N_{S}}\log_{2}\underset{\triangleq\gamma\quad\quad\quad\quad}{\left(1+\underbrace{\frac{1}{N_{S}\sigma_{n}^{2}}}\left[\boldsymbol{\mathbf{\Sigma}}_{E,i}^{2}[k]\right]_{n_{s},n_{s}}\right)}\\
 & \qquad\qquad=\sum_{n_{s}=1}^{N_{S}}\log_{2}\left(1+\gamma\left[\boldsymbol{\mathbf{\Sigma}}_{E,i}^{2}[k]\right]_{n_{s},n_{s}}\right).
\end{alignedat}
\label{eq: I_max}
\end{equation}
It is simpler to find the key parameter of the hybrid beamforming
gain in the high and low SNR regimes. At low SNR ($\gamma\rightarrow0$),
using the fact that $\log(1+\gamma x)\approx\gamma x$ as $\gamma\rightarrow0$,
the achievable data rate in (\ref{eq: I_max}) can be approximated
by
\begin{equation}
\begin{alignedat}{1}I\left(\overline{\mathbf{F}}_{P,i_{f}},\overline{\mathbf{W}}_{P,i_{w}},\overline{\mathbf{F}}_{B,i}[k],\overline{\mathbf{W}}_{B,i}[k]\right) & \overset{\gamma\rightarrow0}{\approx}\gamma\sum_{n_{s}=1}^{N_{S}}\left[\boldsymbol{\mathbf{\Sigma}}_{E,i}^{2}[k]\right]_{n_{s},n_{s}}\\
 & \ \propto\sum_{n_{s}=1}^{N_{S}}\left[\boldsymbol{\mathbf{\Sigma}}_{E,i}^{2}[k]\right]_{n_{s},n_{s}}\\
 & \ \overset{(a)}{\leq}\left\Vert \mathbf{H}_{E,i}[k]\right\Vert _{F}^{2}
\end{alignedat}
\label{eq: Frob_norm}
\end{equation}
with equality in $(a)$ iff $N_{RF}=N_{S}$. For the case of $N_{RF}>N_{S}$,
$\left\Vert \mathbf{H}_{E,i}[k]\right\Vert _{F}^{2}$ corresponds
to the sum of all $N_{RF}$ (instead of only the $N_{S}$ strongest)
eigenvalues of $\mathbf{H}_{E,i}[k]\mathbf{H}_{E,i}^{H}[k]$. Assuming
that the sum of the weaker $N_{RF}-N_{S}$ eigenvalues of $\mathbf{H}_{E,i}[k]\mathbf{H}_{E,i}^{H}[k]$
is small, the approximation of $\sum_{n_{s}=1}^{N_{S}}\left[\boldsymbol{\mathbf{\Sigma}}_{E,i}^{2}[k]\right]_{n_{s},n_{s}}$
by $\left\Vert \mathbf{H}_{E,i}[k]\right\Vert _{F}^{2}$ seems to
be valid for most cases of interest. 

On the other hand, in the high SNR regime ($\gamma\rightarrow\infty$),
using $\log(1+\gamma x)\approx\log\gamma x$ as $\gamma\rightarrow\infty$,
the achievable data rate in (\ref{eq: I_max}) is approximated by
\begin{equation}
\begin{alignedat}{1} & I\left(\overline{\mathbf{F}}_{P,i_{f}},\overline{\mathbf{W}}_{P,i_{w}},\overline{\mathbf{F}}_{B,i}[k],\overline{\mathbf{W}}_{B,i}[k]\right)\\
 & \qquad\qquad\qquad\negthickspace\overset{\gamma\rightarrow\infty}{\approx}\!\sum_{n_{s}=1}^{N_{S}}\log_{2}\left(\gamma\left[\boldsymbol{\mathbf{\Sigma}}_{E,i}^{2}[k]\right]_{n_{s},n_{s}}\right)\\
 & \qquad\qquad\qquad\,\,=\log_{2}\left(\gamma^{N_{S}}\right)+\log_{2}\left(\prod_{n_{s}=1}^{N_{S}}\left[\boldsymbol{\mathbf{\Sigma}}_{E,i}^{2}[k]\right]_{n_{s},n_{s}}\right)\\
 & \qquad\qquad\qquad\,\,\overset{(b)}{=}\log_{2}\left(\gamma^{N_{S}}\right)+\log_{2}\left(\left|\det\left(\mathbf{H}_{E,i}[k]\right)\right|^{2}\right),
\end{alignedat}
\label{eq: determinant}
\end{equation}
which holds with equality in $(b)$ if $N_{RF}=N_{S}$. When the number
$CR$ of propagation paths is much larger than the number $N_{RF}$
of RF chains ($CR\gg N_{RF}$), it is reasonable to conclude that
$\det(\mathbf{H}_{E,i}[k])\neq0$, i.e., $\text{rank}(\mathbf{H}_{E,i}[k])=N_{RF}$. 

As we have seen, either the Frobenius norm of the effective channel
matrix or the absolute value of the determinant of the effective channel
matrix acts as the key parameter for the system throughput. The
discussion focuses on the high and low SNR regimes, and we will provide
more details on approximation error in the numerical results.

\subsection{Hybrid Beamforming Based on Implicit CSI}

In this subsection, we will introduce how to use the coupling coefficients
(or implicit CSI) to obtain the effective channel matrix $\mathbf{H}_{E,i}[k]$.
Once we have $\mathbf{H}_{E,i}[k]$, the solution of the hybrid beamforming
problem can be efficiently found by using the key parameters. First,
let us show the effective channel presented in (\ref{eq:H_E}) again
and approximate the elements of the matrix $\overline{\mathbf{W}}_{P,i_{w}}^{H}\mathbf{H}[k]\overline{\mathbf{F}}_{P,i_{f}}$
by the coupling coefficients as the following equations

\vspace{-0.2cm}

{\small{}
\begin{equation}
\begin{alignedat}{1}\mathbf{H}_{E,i}[k] & \overset{(\ref{eq:H_E})}{=}(\overline{\mathbf{W}}_{P,i_{w}}^{H}\overline{\mathbf{W}}_{P,i_{w}})^{-0.5}\underset{\approx\mathbf{Y}_{i}[k]}{\underbrace{\overline{\mathbf{W}}_{P,i_{w}}^{H}\mathbf{H}[k]\overline{\mathbf{F}}_{P,i_{f}}}}(\overline{\mathbf{F}}_{P,i_{f}}^{H}\overline{\mathbf{F}}_{P,i_{f}})^{-0.5}\\
 & \:\,\approx(\overline{\mathbf{W}}_{P,i_{w}}^{H}\overline{\mathbf{W}}_{P,i_{w}})^{-0.5}\mathbf{Y}_{i}[k](\overline{\mathbf{F}}_{P,i_{f}}^{H}\overline{\mathbf{F}}_{P,i_{f}})^{-0.5}\\
 & \:\,\triangleq\hat{\mathbf{H}}_{E,i}[k],
\end{alignedat}
\label{eq: H_E_2}
\end{equation}
}where the elements of $\mathbf{Y}_{i}[k]$ can be collected from
the coupling coefficients \cite{Chiang2017_WSA}. For example, when
$\overline{\mathbf{F}}_{P,i_{f}}=[\tilde{\mathbf{f}}_{1},\cdots,\tilde{\mathbf{f}}_{N_{RF}}]$
and $\overline{\mathbf{W}}_{P,i_{w}}=[\tilde{\mathbf{w}}_{1},\cdots,\tilde{\mathbf{w}}_{N_{RF}}]$,
where $\tilde{\mathbf{f}}_{n_{rf}}$ is the $n_{rf}^{\text{th}}$
column of $\mathcal{F}$ and $\tilde{\mathbf{w}}_{n_{rf}}$ is the
$n_{rf}^{\text{th}}$ column of $\mathcal{W}$, one has
\begin{equation}
\begin{alignedat}{1}\mathbf{Y}_{i}[k] & =\left[\begin{array}{ccc}
y_{1,1}[k] & \cdots & y_{1,N_{RF}}[k]\\
\vdots & \ddots & \vdots\\
y_{N_{RF},1}[k] & \cdots & y_{N_{RF},N_{RF}}[k]
\end{array}\right]\\
 & =\overline{\mathbf{W}}_{P,i_{w}}^{H}\mathbf{H}[k]\overline{\mathbf{F}}_{P,i_{f}}+\mathbf{Z}[k].
\end{alignedat}
\end{equation}
Therefore, given a pair $(\overline{\mathbf{F}}_{P,i_{f}},\overline{\mathbf{W}}_{P,i_{w}})$
selected from $\mathcal{I_{F}}$ and $\mathcal{I_{W}}$, we can rapidly
obtain the approximation of $\mathbf{H}_{E,i}[k]$, denoted by $\hat{\mathbf{H}}_{E,i}[k]$
in (\ref{eq: H_E_2}).

In brief, the proposed solution can be stated as follows: first obtain
the candidate sets ($\mathcal{I_{F}}$ and $\mathcal{I_{W}}$) and
the approximation of $\mathbf{H}_{E,i}[k]$ from the observations
(or coupling coefficients) $\{y_{n_{w},n_{f}}[k]\,\forall n_{w},n_{f},k\}$,
and then solve the maximization problem in (\ref{eq: max}), which
can be rewritten as
\begin{equation}
\begin{alignedat}{1}\left(\hat{i}_{f},\hat{i}_{w}\right) & ={\displaystyle \underset{\scriptsize\begin{array}{c}
\overline{\mathbf{F}}_{P,i_{f}}\in\mathcal{I_{F}},\\
\overline{\mathbf{W}}_{P,i_{w}}\in\mathcal{I_{W}}
\end{array}}{\arg\,\max}}\,I_{LM}(\overline{\mathbf{F}}_{P,i_{f}},\overline{\mathbf{W}}_{P,i_{w}})\\
 & \approx{\displaystyle \underset{\scriptsize\begin{array}{c}
\overline{\mathbf{F}}_{P,i_{f}}\in\mathcal{I_{F}},\\
\overline{\mathbf{W}}_{P,i_{w}}\in\mathcal{I_{W}},\\
i=(i_{f}-1)I_{W}+i_{w}
\end{array}}{\arg\,\max}}\sum_{k=0}^{K-1}f\left(\hat{\mathbf{H}}_{E,i}[k]\right),
\end{alignedat}
\label{eq: ABS}
\end{equation}
where $f(\hat{\mathbf{H}}_{E,i}[k])$ denotes the analog beam selection
criterion using (\ref{eq: I_max}), (\ref{eq: Frob_norm}), or $\left|\det\left(\mathbf{H}_{E,i}[k]\right)\right|^{2}$
in (\ref{eq: determinant}), with the argument $\hat{\mathbf{H}}_{E,i}[k]$
($\hat{\mathbf{H}}_{E,i}[k]\overset{\text{SVD}}{=}\hat{\mathbf{U}}_{E,i}[k]\hat{\boldsymbol{\mathbf{\Sigma}}}_{E,i}[k]\hat{\mathbf{V}}_{E,i}^{H}[k]$),
given by

\vspace{-0.2cm}{\small{}
\begin{flalign}
 & f\left(\hat{\mathbf{H}}_{E,i}[k]\right)\nonumber \\
 & =\begin{cases}
\sum_{n_{s}=1}^{N_{S}}\log_{2}\left(1+\gamma\left[\hat{\boldsymbol{\mathbf{\Sigma}}}_{E,i}^{2}[k]\right]_{n_{s},n_{s}}\right), & \text{w/o approx.}\\
\left\Vert \hat{\mathbf{H}}_{E,i}[k]\right\Vert _{F}^{2}, & \text{w/ approx. as }\gamma\rightarrow0\\
\left|\det\left(\hat{\mathbf{H}}_{E,i}[k]\right)\right|^{2}, & \text{w/ approx. as }\gamma\rightarrow\infty
\end{cases}\label{eq: selection_criterion}
\end{flalign}
}Next, according to the selected index pair $(\hat{i}_{f},\hat{i}_{w})$,
the selected analog and corresponding digital beamforming matrices
are given by
\begin{equation}
\begin{alignedat}{1}\hat{\mathbf{F}}_{P} & =\overline{\mathbf{F}}_{P,\hat{i}_{f}},\\
\hat{\mathbf{W}}_{P} & =\overline{\mathbf{W}}_{P,\hat{i}_{w}},\\
\hat{\mathbf{F}}_{B}[k] & =(\hat{\mathbf{F}}_{P}^{H}\hat{\mathbf{F}}_{P})^{-0.5}\left[\hat{\mathbf{V}}_{E,\hat{i}}[k]\right]_{:,1:N_{S}},\\
\hat{\mathbf{W}}_{B}[k] & =(\hat{\mathbf{W}}_{P}^{H}\hat{\mathbf{W}}_{P})^{-0.5}\left[\hat{\mathbf{U}}_{E,\hat{i}}[k]\right]_{:,1:N_{S}},
\end{alignedat}
\label{eq: DBF-1}
\end{equation}
where $\hat{i}=(\hat{i}_{f}-1)I_{W}+\hat{i}_{w}$ and $\hat{\mathbf{U}}_{E,\hat{i}}[k]\hat{\boldsymbol{\mathbf{\Sigma}}}_{E,\hat{i}}[k]\hat{\mathbf{V}}_{E,\hat{i}}^{H}[k]=\text{SVD}(\hat{\mathbf{H}}_{E,\hat{i}}[k])$.

The pseudocode of the proposed hybrid beamforming algorithm based
on implicit CSI is shown in \textbf{Algorithm 1}. The advantages of
the proposed algorithm are: (1) channel estimation for large antenna
arrays can be omitted, and (2) even though the set sizes of $\mathcal{I_{F}}$
and $\mathcal{I_{W}}$ are large, the computational overhead is minor.
At low SNR, we just need to calculate the Frobenius norm of the effective
channel matrices, whose elements can be easily obtained from the observations
$\{y_{n_{w},n_{f}}[k]\,\forall n_{w},n_{f},k\}$.

In (\ref{eq: I_max}), we simply assume that the transmit power is
equally allocated to $N_{S}$ data streams to facilitate the process
of finding the best value of the key parameter. Once we find the analog
and digital beamforming matrices, the global maximum throughput can
be further improved by optimizing the power allocation (i.e., by a
water-filling power allocation scheme \cite{Telatar1999}) for $N_{S}$
data streams according to the effective channel condition.
\begin{figure}[t]
\textbf{\small{}\hspace{-0.3cm}}%
\begin{tabular}{|cl|}
\hline 
\multicolumn{2}{|l|}{\textbf{\small{}Algorithm 1: Hybrid beamforming based on implicit
CSI}}\tabularnewline
\hline 
\multicolumn{2}{|l|}{\textbf{\small{}Input: }{\small{}$\{y_{n_{w},n_{f}}[k]\,\forall n_{w},n_{f},k\}$}}\tabularnewline
\multicolumn{2}{|l|}{\textbf{\small{}Output: }{\small{}$\hat{\mathbf{F}}_{P}$, $\hat{\mathbf{W}}_{P}$,
$(\hat{\mathbf{F}}_{B}[k],\hat{\mathbf{W}}_{B}[k])\,\forall k$}}\tabularnewline
{\small{}1. } & \textbf{\small{}Part I \textemdash{} Initial analog beam selection}\tabularnewline
{\small{}2. } & {\small{}Given $\{y_{n_{w},n_{f}}[k]\,\forall n_{w},n_{f},k\}$, select
$M$ analog beam pairs}\tabularnewline
 & {\small{}$(\hat{\mathbf{f}}_{m},\hat{\mathbf{w}}_{m})$, where $m=1,\cdots,M$,
by using (\ref{eq: ABF}).}\tabularnewline
{\small{}3. } & {\small{}Generate two candidate sets $\mathcal{I_{F}}$ and $\mathcal{I_{W}}$
based on $\{\hat{\mathbf{f}}_{m}\,\forall m\}$ }\tabularnewline
 & {\small{}and $\{\hat{\mathbf{w}}_{m}\,\forall m\}$, respectively.}\tabularnewline
{\small{}4. } & \textbf{\small{}Part II \textemdash{} Analog beam selection by different
selection}\tabularnewline
 & \textbf{\small{}criteria }\tabularnewline
{\small{}5. } & {\small{}$\hat{\mathbf{H}}_{E,i}[k]=(\overline{\mathbf{W}}_{P,i_{w}}^{H}\overline{\mathbf{W}}_{P,i_{w}})^{-0.5}\mathbf{Y}_{i}[k](\overline{\mathbf{F}}_{P,i_{f}}^{H}\overline{\mathbf{F}}_{P,i_{f}})^{-0.5}$, }\tabularnewline
 & {\small{}where $\overline{\mathbf{F}}_{P,i_{f}}\in\mathcal{I_{F}}$,
$\overline{\mathbf{W}}_{P,i_{w}}\in\mathcal{I_{W}}$, and the entries
of $\mathbf{Y}_{i}[k]$}\tabularnewline
 & {\small{}are collected from $\{y_{n_{w},n_{f}}[k]\,\forall n_{w},n_{f}\}.$}\tabularnewline
{\small{}6. } & {\small{}$(\hat{i}_{f},\hat{i}_{w})={\displaystyle \underset{\scriptsize\begin{array}{c}
i=(i_{f}-1)I_{W}+i_{w}\end{array}}{\arg\,\max}\sum_{k=0}^{K-1}f(\hat{\mathbf{H}}_{E,i}[k]),}$}\tabularnewline
 & {\small{}where $f(\hat{\mathbf{H}}_{E,i}[k])$ is given by (\ref{eq: selection_criterion}). }\tabularnewline
{\small{}7. } & {\small{}\uline{Output}}{\small{}: $\hat{\mathbf{F}}_{P}=\overline{\mathbf{F}}_{P,\hat{i}_{f}}$
and $\hat{\mathbf{W}}_{P}=\overline{\mathbf{W}}_{P,\hat{i}_{w}}$.}\tabularnewline
{\small{}8. } & \textbf{\small{}Part III \textemdash{} Corresponding optimal digital
beamforming}\tabularnewline
{\small{}9. } & {\small{}$\hat{\mathbf{U}}_{E,\hat{i}}[k]\hat{\boldsymbol{\mathbf{\Sigma}}}_{E,\hat{i}}[k]\hat{\mathbf{V}}_{E,\hat{i}}^{H}[k]=\text{SVD}(\hat{\mathbf{H}}_{E,\hat{i}}[k])$,}\tabularnewline
 & {\small{}where $\hat{i}=(\hat{i}_{f}-1)I_{W}+\hat{i}_{w}$.}\tabularnewline
{\small{}10. } & {\small{}\uline{Output}}{\small{}: $\begin{cases}
\hat{\mathbf{F}}_{B}[k]=(\hat{\mathbf{F}}_{P}^{H}\hat{\mathbf{F}}_{P})^{-0.5}\left[\hat{\mathbf{V}}_{E,\hat{i}}[k]\right]_{:,1:N_{S}}\\
\hat{\mathbf{W}}_{B}[k]=(\hat{\mathbf{W}}_{P}^{H}\hat{\mathbf{W}}_{P})^{-0.5}\left[\hat{\mathbf{U}}_{E,\hat{i}}[k]\right]_{:,1:N_{S}}
\end{cases}$}\tabularnewline
\hline 
\end{tabular}
\end{figure}

\section{Analysis of the Proposed Hybrid Beamforming Algorithm\label{sec:Analysis-of-the}}

In the section, we focus on the statistical analysis of using the
Frobenius norm of the effective channel as the key parameter at low
SNR. Starting from (\ref{eq: H_E_2}), $\hat{\mathbf{H}}_{E,i}[k]$
can be expressed as a noisy version of the true effective channel
as

\vspace{-0.2cm}{\small{}
\begin{equation}
\begin{alignedat}{1}\hat{\mathbf{H}}_{E,i}[k] & \negthickspace\overset{(\ref{eq: H_E_2})}{=}(\overline{\mathbf{W}}_{P,i_{w}}^{H}\overline{\mathbf{W}}_{P,i_{w}})^{-0.5}\mathbf{Y}_{i}[k](\overline{\mathbf{F}}_{P,i_{f}}^{H}\overline{\mathbf{F}}_{P,i_{f}})^{-0.5}\\
 & =\underset{\mathbf{H}_{E,i}[k]}{\underbrace{(\overline{\mathbf{W}}_{P,i_{w}}^{H}\overline{\mathbf{W}}_{P,i_{w}})^{-0.5}\overline{\mathbf{W}}_{P,i_{w}}^{H}\mathbf{H}[k]\overline{\mathbf{F}}_{P,i_{f}}(\overline{\mathbf{F}}_{P,i_{f}}^{H}\overline{\mathbf{F}}_{P,i_{f}})^{-0.5}}}\\
 & \quad\,+\underset{\mathbf{Z}_{E,i}[k]}{\underbrace{(\overline{\mathbf{W}}_{P,i_{w}}^{H}\overline{\mathbf{W}}_{P,i_{w}})^{-0.5}\mathbf{Z}[k](\overline{\mathbf{F}}_{P,i_{f}}^{H}\overline{\mathbf{F}}_{P,i_{f}})^{-0.5}}}\\
 & =\mathbf{H}_{E,i}[k]+\mathbf{Z}_{E,i}[k],
\end{alignedat}
\label{eq: H_E_hat}
\end{equation}
}where the multivariate distribution of the $N_{RF}^{2}$-dimensional
random vector $\text{vec}(\mathbf{Z}_{E,i}[k])$ can be written as
(see Appendix \ref{sec:Derivation-of-the})\vspace{-0.2cm}
\begin{equation}
\begin{alignedat}{1} & \text{vec}(\mathbf{Z}_{E,i}[k])\\
 & \sim\mathcal{CN}\left(\boldsymbol{0}_{N_{RF}^{2}\times1},\sigma_{n}^{2}\left((\overline{\mathbf{F}}_{P,i_{f}}^{T}\overline{\mathbf{F}}_{P,i_{f}}^{*})^{-1}\otimes(\overline{\mathbf{W}}_{P,i_{w}}^{H}\overline{\mathbf{W}}_{P,i_{w}})^{-1}\right)\right).
\end{alignedat}
\label{eq: vec_Z}
\end{equation}

From (\ref{eq: H_E_hat}), we have $||\hat{\mathbf{H}}_{E,i}[k]||_{F}^{2}$
given by
\begin{multline}
\underset{\text{estimate of the key parameter}}{\underbrace{\left\Vert \hat{\mathbf{H}}_{E,i}[k]\right\Vert _{F}^{2}}}=\underset{\text{true value of the key parameter}}{\underbrace{\left\Vert \mathbf{H}_{E,i}[k]\right\Vert _{F}^{2}}}\\
+\underset{\text{noise}}{\underbrace{\left\Vert \mathbf{Z}_{E,i}[k]\right\Vert _{F}^{2}+2\cdot\mathfrak{R}\left(\text{tr}(\mathbf{H}_{E,i}^{H}[k]\mathbf{Z}_{E,i}[k])\right)}}.\label{eq: Forb_H_E_hat}
\end{multline}
To analyze the noise effect, we introduce the quantities $U$ and
$V$, conditional on a channel state $\mathbf{H}'[k]$ and an analog
beamforming pair $(\overline{\mathbf{F}}_{P,i_{f}},\overline{\mathbf{W}}_{P,i_{w}})$,
given by
\begin{flalign}
U & =\left\Vert \mathbf{Z}_{E,i}[k]\right\Vert _{F}^{2},\label{eq: U}\\
V & =2\cdot\mathfrak{R}\left(\text{tr}\left((\mathbf{H}_{E,i}'[k])^{H}\mathbf{Z}_{E,i}[k]\right)\right),\label{eq: V}
\end{flalign}
where $\mathbf{H}_{E,i}'[k]=\overline{\mathbf{W}}_{P,i_{w}}^{H}\mathbf{H}'[k]\overline{\mathbf{F}}_{P,i_{f}}$,
and then pursue the analysis of $U$ and $V$ for orthogonal and non-orthogonal
codebooks. 

\subsection{Orthogonal codebooks\label{subsec:Orthogonal-codebooks}}

When the columns of $\overline{\mathbf{F}}_{P,i_{f}}$ and the columns
of $\overline{\mathbf{W}}_{P,i_{w}}$ are mutually orthogonal respectively,
from (\ref{eq: vec_Z}) we know that the elements of $\text{vec}(\mathbf{Z}_{E,i}[k])$
have the same normal distribution with mean zero and variance $\sigma_{n}^{2}$,
$\text{vec}(\mathbf{Z}_{E,i}[k])\sim\mathcal{CN}(\boldsymbol{0}_{N_{RF}^{2}\times1},\sigma_{n}^{2}\mathbf{I}_{N_{RF}^{2}})$.
Therefore, $U$ is the sum of the absolute squares of $N_{RF}^{2}$
i.i.d. Gaussian random variables, which follows a Gamma distribution
with shape parameter $N_{RF}^{2}$ and scale parameter $\sigma_{n}^{2}$:
\begin{equation}
\begin{alignedat}{1}U & =\sum_{i=1}^{N_{RF}}\sum_{j=1}^{N_{RF}}\underset{\sim\Gamma\left(\frac{1}{2},\sigma_{n}^{2}\right)}{\underbrace{\mathfrak{R}\left(\left[\mathbf{Z}_{E,i}[k]\right]_{i,j}\right)^{2}}}+\underset{\sim\Gamma\left(\frac{1}{2},\sigma_{n}^{2}\right)}{\underbrace{\mathfrak{I}\left(\left[\mathbf{Z}_{E,i}[k]\right]_{i,j}\right)^{2}}}\\
 & \sim\Gamma(N_{RF}^{2},\sigma_{n}^{2}).
\end{alignedat}
\label{eq: U_orthogonal}
\end{equation}
In addition, $V$ is normally distributed with mean zero and variance
$2\sigma_{n}^{2}\left\Vert \mathbf{H}_{E,i}'[k]\right\Vert _{F}^{2}$.

\subsection{Non-orthogonal codebooks\label{subsec:Non-orthogonal-Codebooks}}

When the columns of $\overline{\mathbf{F}}_{P,i_{f}}$ or the columns
of $\overline{\mathbf{W}}_{P,i_{w}}$ are not mutually orthogonal,
the elements of $\text{vec}(\mathbf{Z}_{E,i}[k])$ in (\ref{eq: vec_Z})
are not i.i.d. anymore. In this case, there are no closed-form expressions
for the probability distributions of $U$ and $V$. Accordingly, we
only derive and state $\text{E}[U]$, $\text{Var}(U)$, and $\text{E}[V]$
in this section. These are given by (see Appendix \ref{sec:Derivation-of-})
\begin{equation}
\text{E}[U]=\sigma_{n}^{2}\cdot\text{tr}\left((\overline{\mathbf{F}}_{P,i_{f}}^{T}\overline{\mathbf{F}}_{P,i_{f}}^{*})^{-1}\right)\text{tr}\left((\overline{\mathbf{W}}_{P,i_{w}}^{H}\overline{\mathbf{W}}_{P,i_{w}})^{-1}\right),
\end{equation}
and
\begin{equation}
\text{Var}(U)=\text{tr}(\boldsymbol{\Psi}\mathbf{R}_{z_{V}})-\text{E}[U]^{2},
\end{equation}
where
\begin{flalign}
\boldsymbol{\Psi} & =\left((\overline{\mathbf{F}}_{P,i_{f}}^{T}\overline{\mathbf{F}}_{P,i_{f}}^{*})^{-1}\otimes(\overline{\mathbf{W}}_{P,i_{w}}^{H}\overline{\mathbf{W}}_{P,i_{w}})^{-1}\right)\nonumber \\
 & \quad\,\otimes\left((\overline{\mathbf{F}}_{P,i_{f}}^{T}\overline{\mathbf{F}}_{P,i_{f}}^{*})^{-1}\otimes(\overline{\mathbf{W}}_{P,i_{w}}^{H}\overline{\mathbf{W}}_{P,i_{w}})^{-1}\right),
\end{flalign}

\begin{flalign}
\mathbf{R}_{z_{V}} & =\text{E}[\underset{\mathbf{z}_{V}[k]\in\mathbb{C}^{N_{RF}^{4}\times1}}{\underbrace{\left(\text{vec}(\mathbf{Z}[k])\otimes\text{vec}(\mathbf{Z}[k])\right)}}\underset{\mathbf{z}_{V}^{H}[k]}{\underbrace{\left(\text{vec}(\mathbf{Z}[k])\otimes\text{vec}(\mathbf{Z}[k])\right)^{H}}}],
\end{flalign}
and $\text{E}[V]=0$. Unfortunately, we did not find a closed-form
expression for $\text{Var}(V)$. From the analysis results, it is
clear that when we use non-orthogonal codebooks, the distributions
of $U$ and $V$ vary with different candidates for analog beamforming
matrices. This implies that $||\hat{\mathbf{H}}_{E,i}[k]||_{F}^{2}$
in (\ref{eq: Forb_H_E_hat}) may become unreliable because of the
non-i.i.d. noise signals.

\section{Simulation Results\label{sec:Simulation-Results}}

The system parameters used in the simulations are listed below. In
addition, the SNR in linear scale is given by $\text{SNR}=\frac{\rho}{N_{S}\sigma_{n}^{2}}$,
where $\rho$ is the average received power.
\begin{figure}[t]
\centering{}%
\begin{tabular}{|cl|}
\hline 
\multicolumn{2}{|l|}{\textbf{\small{}Algorithm 2: Hybrid beamforming based on explicit
CSI}}\tabularnewline
\hline 
\multicolumn{2}{|l|}{\textbf{\small{}Input: }{\small{}$\{\mathbf{V}[k]\:\forall k\}$}}\tabularnewline
\multicolumn{2}{|l|}{\textbf{\small{}Output: }{\small{}$\check{\mathbf{F}}_{P}$, $\check{\mathbf{F}}_{B}[k]\:\forall k$}}\tabularnewline
{\small{}1.} & {\small{}$\check{\mathbf{F}}_{P}=\text{empty matrix}$}\tabularnewline
{\small{}2.} & {\small{}$\mathbf{V}_{R}[k]=\left[\mathbf{V}[k]\right]_{:,1:N_{S}}$ }\tabularnewline
{\small{}3.} & {\small{}for $n_{rf}=1,\cdots,N_{RF}$}\tabularnewline
{\small{}4.} & {\small{}$\qquad$$\check{\mathbf{f}}_{P,n_{rf}}={\displaystyle \underset{\scriptsize\begin{array}{c}
\tilde{\mathbf{f}}_{n_{f}}\in\mathcal{F}\end{array}}{\arg\,\max}\sum_{k=0}^{K-1}\left\Vert \tilde{\mathbf{f}}_{n_{f}}^{H}\mathbf{V}_{R}[k]\right\Vert _{F}^{2}}$}\tabularnewline
{\small{}5.} & {\small{}$\qquad$$\check{\mathbf{F}}_{P}=[\check{\mathbf{F}}_{P}\,|\,\check{\mathbf{f}}_{P,n_{rf}}]$}\tabularnewline
{\small{}6.} & {\small{}$\qquad$$\mathbf{V}_{R}[k]=(\mathbf{I}_{N_{T}}-\check{\mathbf{F}}_{P}(\check{\mathbf{F}}_{P}^{H}\check{\mathbf{F}}_{P})^{-1}\check{\mathbf{F}}_{P}^{H})\left[\mathbf{V}[k]\right]_{:,1:N_{S}}$}\tabularnewline
{\small{}7.} & {\small{}$\qquad$$\mathbf{V}_{R}[k]=\frac{\mathbf{V}_{R}[k]}{\left\Vert \mathbf{V}_{R}[k]\right\Vert _{F}}$}\tabularnewline
{\small{}8.} & {\small{}end}\tabularnewline
{\small{}9.} & {\small{}$\check{\mathbf{F}}_{B}[k]=(\check{\mathbf{F}}_{P}^{H}\check{\mathbf{F}}_{P})^{-1}\check{\mathbf{F}}_{P}^{H}\left[\mathbf{V}[k]\right]_{:,1:N_{S}}$}\tabularnewline
{\small{}10.} & {\small{}$\check{\mathbf{F}}_{B}[k]=\sqrt{N_{S}}\cdot\frac{\check{\mathbf{F}}_{B}[k]}{\left\Vert \check{\mathbf{F}}_{P}\check{\mathbf{F}}_{B}[k]\right\Vert _{F}}$}\tabularnewline
\hline 
\end{tabular}
\end{figure}
\\
\\
\begin{tabular}{ll}
Number of antennas & $N_{T}=N_{R}=32$\tabularnewline
Number of RF chains & $N_{RF}=2$ \tabularnewline
Number of data streams & $N_{S}=2$ \tabularnewline
Length of a training sequence & $K=512$\tabularnewline
Number of clusters & $C=5$ ($1$ LoS and $4$ NLoS\tabularnewline
 & clusters)\tabularnewline
Number of rays per cluster & $R=8$\tabularnewline
\end{tabular}\\

We chose the work in \cite{Ayach2012} that implements hybrid beamforming
based on \textit{explicit} \textit{CSI} as a reference method for
comparison and extended it from single carrier to multiple carriers.
In the reference method, given the channel matrices, $\mathbf{H}[k]\overset{\text{SVD}}{=}\mathbf{U}[k]\boldsymbol{\Sigma}[k]\mathbf{V}^{H}[k]\:\forall k$,
the goal of the precoder design is to minimize the sum of the squared
Frobenius norms of the errors between the right singular vectors and
the precoder across all subcarriers:

{\small{}\vspace{-0.3cm}
\begin{equation}
\begin{gathered}(\check{\mathbf{F}}_{P},\check{\mathbf{F}}_{B}[k]\:\forall k)={\displaystyle \underset{\mathbf{F}_{P},\mathbf{F}_{B}[k]\,\forall k}{\arg\text{ }\min}}\sum_{k=0}^{K-1}\left\Vert \left[\mathbf{V}[k]\right]_{:,1:N_{S}}-\mathbf{F}_{P}\mathbf{F}_{B}[k]\right\Vert _{F}^{2},\\
\text{s.t. }\begin{cases}
\mathbf{f}_{P,n_{rf}}\in\mathcal{F}\;\forall n_{rf},\\
\left\Vert \mathbf{F}_{P}\mathbf{F}_{B}[k]\right\Vert _{F}^{2}=N_{S}\:\forall k.
\end{cases}
\end{gathered}
\label{eq: ref_F}
\end{equation}
}The problem can be solved by the OMP algorithm \cite{Cai2011} and
the pseudocode is given in \textbf{Algorithm 2}. Similarly, given
$\left[\mathbf{U}[k]\right]_{:,1:N_{S}}$, we have the corresponding
solution to the combiner, denoted by $(\check{\mathbf{W}}_{P},\check{\mathbf{W}}_{B}[k]\:\forall k)$. 

Comparing \textbf{Algorithm 1} with \textbf{Algorithm 2}, we know
that the first algorithm uses the received coupling coefficients as
the inputs, while the second uses the singular vectors of the channel
as the inputs. The coupling coefficients are commonly used for channel
estimation \cite{Mendez-Rial2016,Chiang2016_ISWCS}, but in this paper
we use them to directly implement the hybrid beamforming on both sides.
As a result, we can get rid of the overhead of channel estimation.

To clearly present the difference in throughput in the simulation
results, the calculated throughput values are normalized to the throughput
achieved by fully digital beamforming (DBF) given by
\begin{equation}
I_{DBF}=\frac{1}{K}\sum_{k=0}^{K-1}\sum_{n_{s}=1}^{N_{S}}\log_{2}\left(1+\gamma\left[\boldsymbol{\Sigma}^{2}[k]\right]_{n_{s},n_{s}}\right),\label{eq: Fully_DBF}
\end{equation}
where $\gamma=\frac{1}{N_{S}\sigma_{n}^{2}}$ and the diagonal entries
of $\boldsymbol{\Sigma}^{2}[k]$ are the eigenvalues of $\mathbf{H}[k]\mathbf{H}^{H}[k]$.
The data rates achieved by (\ref{eq: Fully_DBF}) used for the normalization
from $\text{SNR}=-20$ dB to $30$ dB (step by $5$ dB) are: $\{0.05,0.14,0.41,1.03,2.17,3.77,5.79,8.17,$\\
$10.91,13.95,17.13\}$ in bit/s/Hz. In what follows, a complete analysis
with respect to three different codebooks, whose coherence values
are $0$, $0.12$, and $0.99$, is provided\footnote{The coherence of a codebook $\mathcal{F}$ is defined as $\max_{i<j}\frac{\left|\tilde{\mathbf{f}}_{i}^{H}\tilde{\mathbf{f}}_{j}\right|}{\left\Vert \tilde{\mathbf{f}}_{i}\right\Vert _{2}\left\Vert \tilde{\mathbf{f}}_{j}\right\Vert _{2}}$
\cite{Candes2011}.}.

\subsection{Orthogonal codebooks}

Assume that the codebooks $\mathcal{F}$ and $\mathcal{W}$ have the
same number $N_{F}=N_{W}=32$ of candidates for the analog beamforming
vectors. When $\mathcal{F}$ and $\mathcal{W}$ are orthogonal codebooks,
the $32$ candidates for the steering spatial frequency are equally
distributed in the spatial frequency domain and the corresponding
steering angles are: $\left\{ \frac{180^{\circ}}{\pi}\cdot\sin^{-1}\left(\frac{\left(n_{f}-16\right)}{16}\right),\,n_{f}=1,\cdots,32\right\} $
\cite{Chiang2016_WOWMOM}.

\begin{figure}[t]
\centering{}\hspace*{-0.3cm}\includegraphics[scale=0.58]{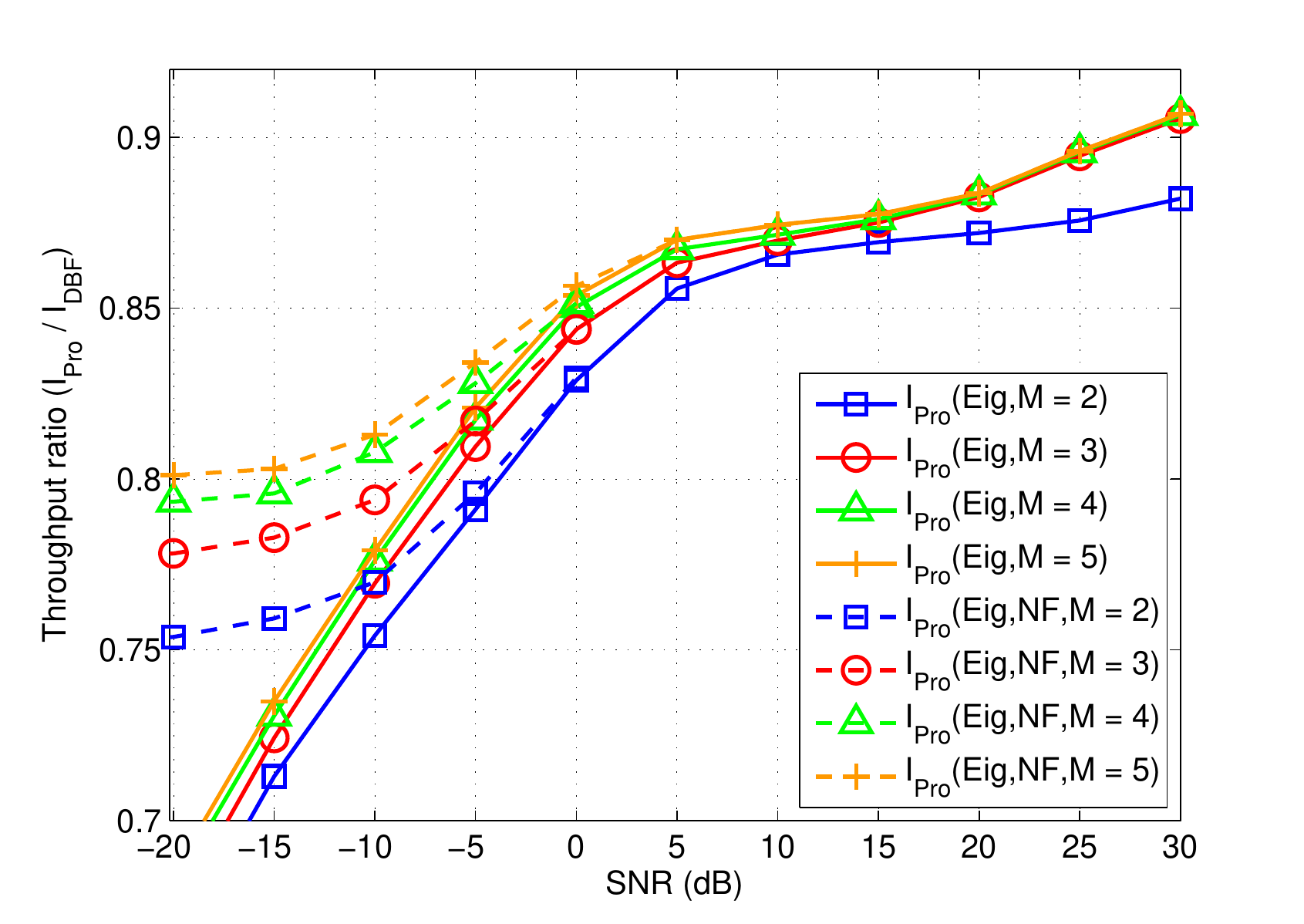}\caption{Achievable throughput, normalized to $I_{DBF}$, by the proposed methods
with the orthogonal codebook and $M=2,3,4,5$ initially selected analog
beam pairs.\label{fig: orthogonal_eigen} }
\end{figure}

In Fig. \ref{fig: orthogonal_eigen}, we evaluate the achievable
data rates with $M=2,3,4,5$ initially selected analog beam pairs
in the proposed method, and more details of these curves are stated
as follows: 
\begin{itemize}
\item \textit{$I_{Pro}(Eig,M=2,3,4,5)$} is calculated by\textbf{
\begin{equation}
I_{Pro}=\frac{1}{K}\sum_{k=0}^{K-1}I(\hat{\mathbf{F}}_{P},\hat{\mathbf{W}}_{P},\hat{\mathbf{F}}_{B}[k],\hat{\mathbf{W}}_{B}[k]),\label{eq: I_pro}
\end{equation}
}where \textbf{$(\hat{\mathbf{F}}_{P},\hat{\mathbf{W}}_{P},\hat{\mathbf{F}}_{B}[k],\hat{\mathbf{W}}_{B}[k])$}
is the output of \textbf{Algorithm 1} with the beam selection criterion
$f(\hat{\mathbf{H}}_{E,i}[k])=\sum_{n_{s}=1}^{N_{S}}\log_{2}(1+\gamma[\hat{\boldsymbol{\mathbf{\Sigma}}}_{E,i}^{2}[k]]_{n_{s},n_{s}})$
in Step 6 in the algorithm. In the phase of initial analog beam selection,
we reserve $M=2,3,4,5$ initially selected analog beam pairs.\\
\item $I_{Pro}(Eig,N\!F,M\!=\!2,3,4,5)$ is calculated by the same way as
$I_{Pro}(Eig,M=2,3,4,5)$ but with noise-free observations. That is
to say, the inputs of \textbf{Algorithm 1}, $\{y_{n_{w},n_{f}}[k]\,\forall n_{w},n_{f},k\}$,
do not take into account the noise effect.
\end{itemize}
As shown in Fig. \ref{fig: orthogonal_eigen}, in the low SNR regime,
the beam selection performance more or less suffers from the noise
effect. When $\text{SNR}>0\text{ dB}$, the noisy observations are
reliable enough to achieve almost the same throughput as that by using
the noise-free observations. In addition, when $M>5$, the achievable
data rates are almost the same as the curves with $M=5$ (although
they are not shown in the figure)\footnote{The number ``$M-N_{RF}$'' can be interpreted as a degree of diversity.
As it is well known from other diversity techniques, the gain in performance
decays more or less quickly with the increasing degrees of diversity.}. For the sake of low complexity, it is not necessary to take more
than $M=5$ candidates into account because most observations are
dominated by noise signals except for those corresponding to the already
selected analog beamforming vectors.

\begin{figure}[t]
\centering{}\hspace*{-0.3cm}\includegraphics[scale=0.58]{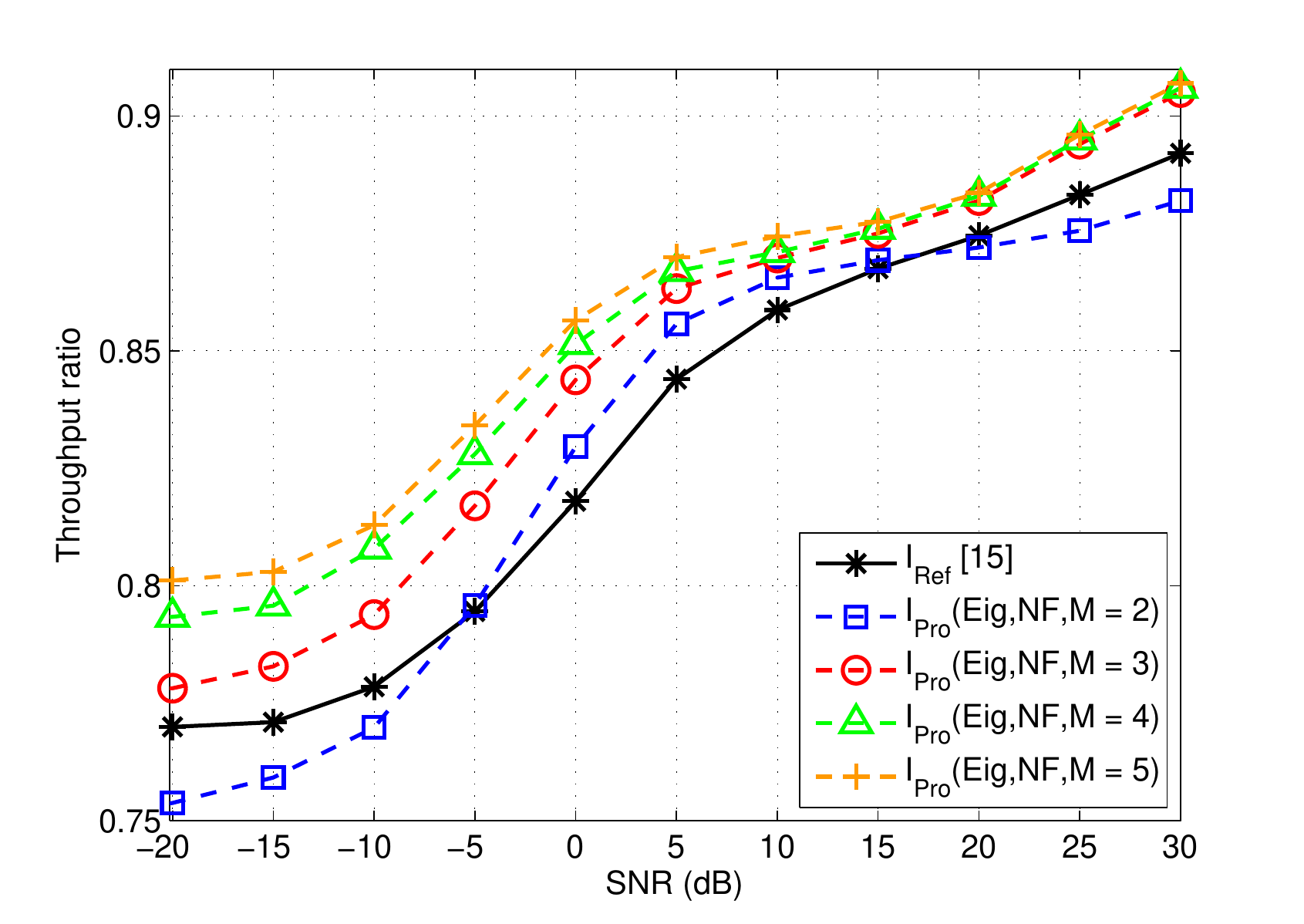}\caption{Comparisons between the reference method and the proposed approach
with noise-free (NF) observations, i.e., the curves $I_{Pro}(Eig,N\!F,M=2,3,4,5)$
in Fig. \ref{fig: orthogonal_eigen}.\label{fig: orthogonal_ref_vs_eigen} }
\end{figure}

Comparisons between the proposed and reference methods are shown
in Fig. \ref{fig: orthogonal_ref_vs_eigen}. To better compare our
approach with the reference method, we choose the curves $I_{Pro}(Eig,N\!F,M=2,3,4,5)$
in Fig. \ref{fig: orthogonal_eigen}, whose inputs are noise-free
observations. Furthermore, the reference curve denoted by \textit{$I_{Ref}$}
is calculated by\textbf{
\begin{equation}
I_{Ref}=\frac{1}{K}\sum_{k=0}^{K-1}I(\check{\mathbf{F}}_{P},\check{\mathbf{W}}_{P},\check{\mathbf{F}}_{B}[k],\check{\mathbf{W}}_{B}[k]),
\end{equation}
}where $(\check{\mathbf{F}}_{P},\check{\mathbf{W}}_{P},\check{\mathbf{F}}_{B}[k],\check{\mathbf{W}}_{B}[k])$
is obtained from \textbf{Algorithm 2} with the inputs $\{\mathbf{V}[k]\:\forall k\}$
and $\{\mathbf{U}[k]\:\forall k\}$. Data rates achieved by the reference
and proposed methods shown in Fig. \ref{fig: orthogonal_ref_vs_eigen}
are normalized to $I_{DBF}$, i.e., $I_{Ref}/I_{DBF}$ and $I_{Pro}/I_{DBF}$.
In the figure, we can find that the curves \textit{$I_{Pro}(Eig,N\!F,M=3,4,5)$
}achieve higher data rates than \textit{$I_{Ref}$}. Although these
two methods use different ways to implement the hybrid beamforming,
we try an explanation based on some assumptions. Assume that these
two schemes find the same $N_{RF}$ analog beam pairs (i.e., $\hat{\mathbf{F}}_{P}=\check{\mathbf{F}}_{P}$
and $\hat{\mathbf{W}}_{P}=\check{\mathbf{W}}_{P}$), which means that
they have the same effective channel $\mathbf{H}_{E}[k]=\hat{\mathbf{W}}_{P}^{H}\mathbf{H}[k]\hat{\mathbf{F}}_{P}=\check{\mathbf{W}}_{P}^{H}\mathbf{H}[k]\check{\mathbf{F}}_{P}$.
In this case, \textbf{Algorithm 1} uses the SVD of $\mathbf{H}_{E}[k]$
to find the solution of digital beamforming matrices. From \cite{Telatar1999},
we know that this solution is optimal. In contrast, the digital beamforming
in \textbf{Algorithm 2} Step 9 uses the least-squares solution, which
is sub-optimal. When we reserve more candidates ($M>N_{RF}$), it
has a high probability that both algorithms find the same $N_{RF}$
analog beam pairs. If so, \textbf{Algorithm 1} theoretically outperforms
\textbf{Algorithm 2}.

\begin{figure}[t]
\centering{}\hspace*{-0.3cm}\includegraphics[scale=0.58]{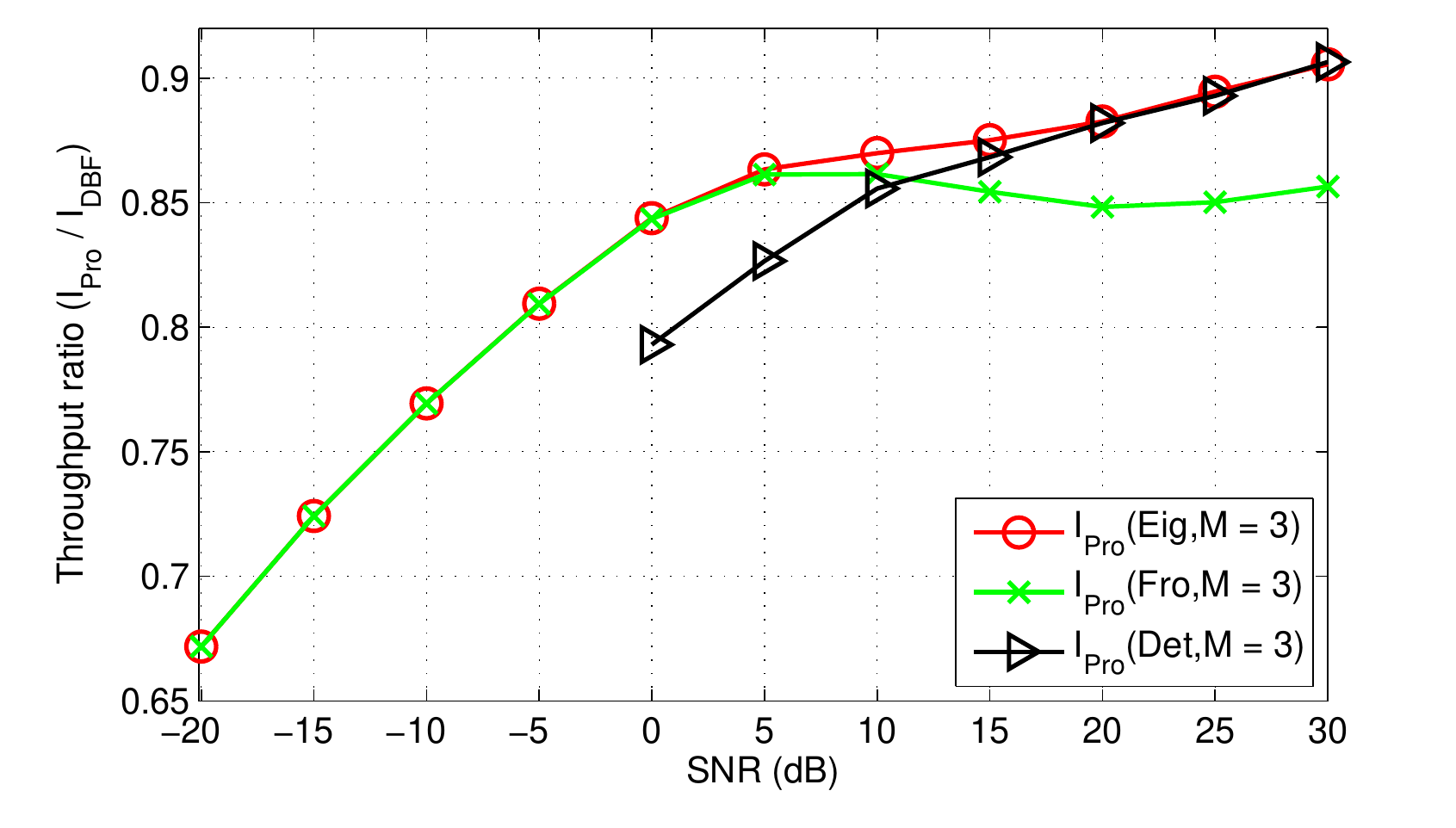}\caption{The curve $I_{Pro}(Eig,M=3)$ shown in Fig. \ref{fig: orthogonal_eigen}
and its approximations achieved by using two different key parameters.\label{fig:orthogonal_cb} }
\end{figure}

Next, approximation results of \textit{$I_{Pro}(Eig,M=3)$} by using
the key parameters are shown in Fig. \ref{fig:orthogonal_cb}, where
\begin{itemize}
\item \textit{$I_{Pro}(Fro,M\!=\!3)$} and \textit{$I_{Pro}(Det,M\!=\!3)$}
use beam selection criteria $f(\hat{\mathbf{H}}_{E,i}[k])=||\hat{\mathbf{H}}_{E,i}[k]||_{F}^{2}$
and $f(\hat{\mathbf{H}}_{E,i}[k])=|\det(\hat{\mathbf{H}}_{E,i}[k])|^{2}$,
respectively, in \textbf{Algorithm 1} Step 6. 
\end{itemize}
When $\text{SNR}<5\text{ dB}$, \textit{$I_{Pro}(Fro,M=3)$} achieves
almost the same throughput as \textit{$I_{Pro}(Eig,M=3)$}. When $5\text{ dB}<\text{SNR}<20\text{ dB}$,
both key parameters cannot perfectly yield the same data rates as
\textit{$I_{Pro}(Eig,M=3)$}, but the relative loss amounts to at
most a few percentages. From Fig. \ref{fig: orthogonal_eigen}, Fig.
\ref{fig: orthogonal_ref_vs_eigen}, and Fig. \ref{fig:orthogonal_cb},
we can find that if the system operates in the SNR range of $0$ to
$5$ dB, the Frobenius norm of the estimated effective channel works
pretty well.

\subsection{Non-orthogonal codebooks}

Now assume that $\mathcal{F}$ and $\mathcal{W}$ are non-orthogonal
codebooks. As mentioned in Section \ref{sec:Analysis-of-the}, if
the columns of $\mathcal{F}$ or $\mathcal{W}$ are not mutually orthogonal,
some highly correlated columns (e.g., $\tfrac{|\tilde{\mathbf{f}}_{i}^{H}\tilde{\mathbf{f}}_{j}|}{||\tilde{\mathbf{f}}_{i}||_{2}||\tilde{\mathbf{f}}_{j}||_{2}}=0.99,i\neq j$)
may make the effective noise level unacceptably large, and the estimated
effective channel becomes unreliable accordingly. Here we use two
non-orthogonal codebooks to characterize the noise effect:
\begin{itemize}
\item The first non-orthogonal codebook has $N_{F}$ $=$ $N_{W}$ $=$
$36$ columns and the corresponding $36$ steering angles are: $\left\{ \frac{180^{\circ}}{\pi}\cdot\sin^{-1}\left(\frac{\left(n_{f}-18\right)}{18}\right),\,n_{f}=1,\cdots,36\right\} $.
The coherence of the codebook is $0.12$ that implies a weakly coherent
codebook . 
\item The second non-orthogonal codebook has larger coherence than the first
one. It has $N_{F}=N_{W}=32$ columns and the corresponding $32$
steering angles are: $\left\{ -90^{o}+\frac{180^{o}\cdot n_{f}}{N_{F}},n_{f}=1,\cdots,32\right\} $.
This codebook design leads to the coherence of $0.99$ that implies
a strongly coherent codebook.
\end{itemize}
\begin{figure}[t]
\centering{}\hspace*{-0.3cm}\includegraphics[scale=0.58]{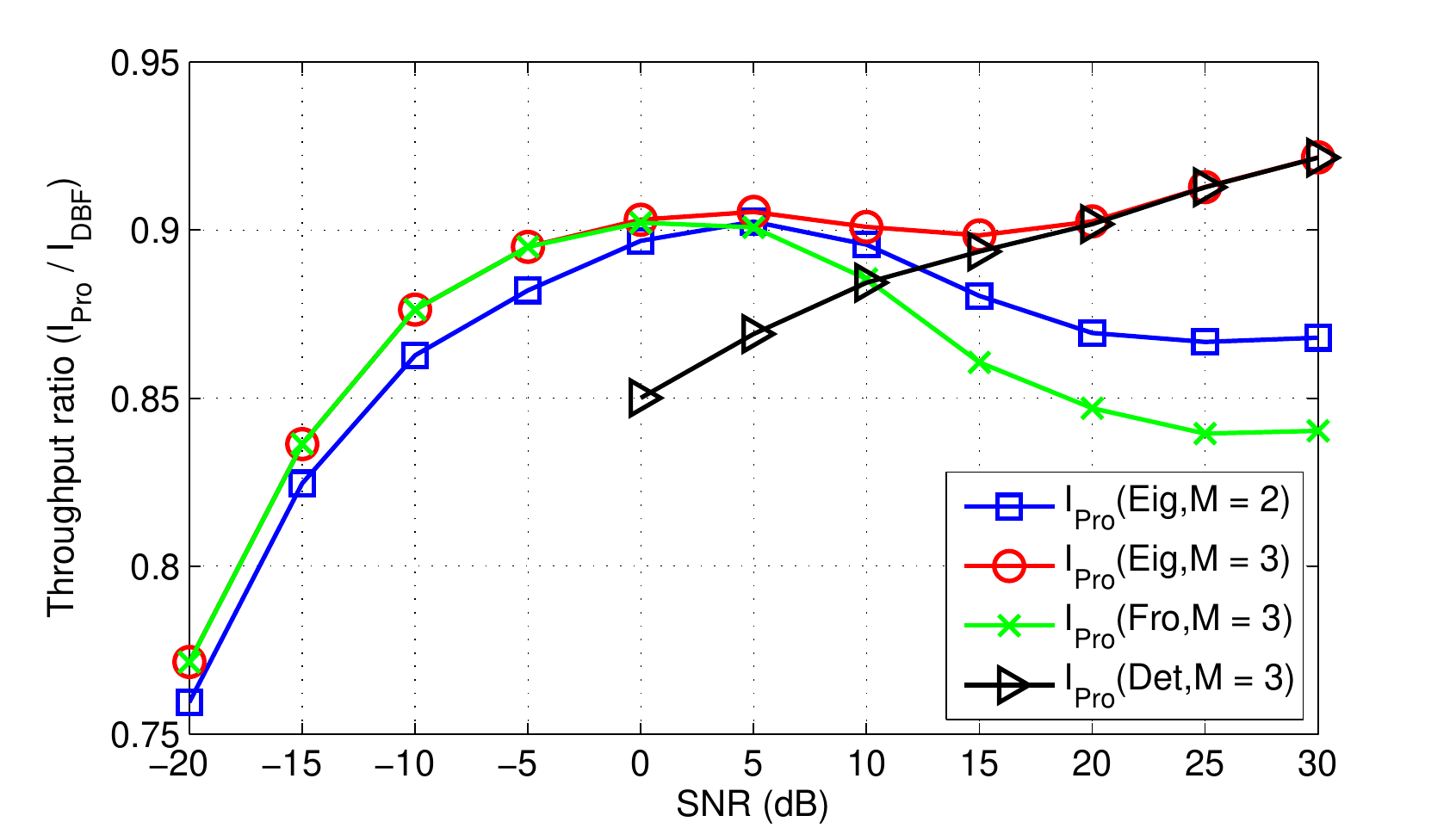}\caption{Achievable throughput, normalized to $I_{DBF}$, by the proposed approach
with the weakly coherent codebook and comparisons between \textit{$I_{Pro}(Eig,M=3)$}
and its approximations.\label{fig: incoherent_cb} }
\end{figure}

In Fig. \ref{fig: incoherent_cb}, when using the weakly coherent
codebook (coherence $=0.12$) at the transmitter and receiver, the
throughput shown in the curve \textit{$I_{Pro}(Eig,M=3)$} can be
further improved compared with \textit{$I_{Pro}(Eig,M=2)$}, which
means that, in (\ref{eq: H_E_hat}), the effect of the correlated
columns of the weakly coherent codebook on $\hat{\mathbf{H}}_{E,i}[k]$
is minor. Also, the approximations of $I_{Pro}(Eig,M=3)$ by using
the key parameters shown in the curves $I_{Pro}(Fro,M=3)$ at low
SNR and $I_{Pro}(Det,M=3)$ at high SNR are quite accurate and only
with small differences in the SNR range of 0 to 20 dB. Compared with
the results shown in Fig. \ref{fig:orthogonal_cb} with the orthogonal
codebook, although the approximations in the case of the weakly coherent
codebook become slightly worse, the achievable throughput overall
becomes better since there are four additional candidates in the weakly
coherent codebook.

\begin{figure}[t]
\centering{}\hspace*{-0.3cm}\includegraphics[scale=0.58]{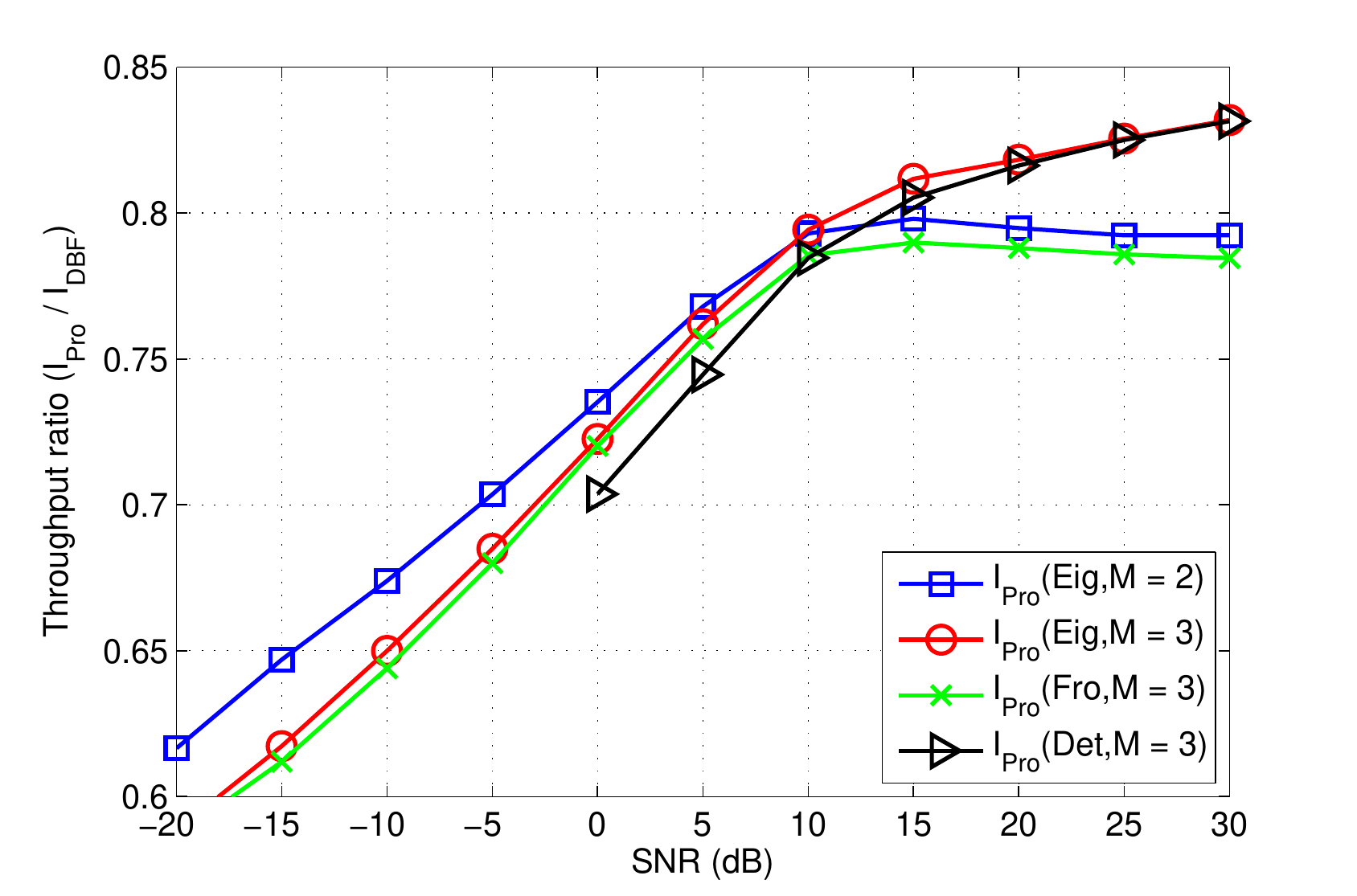}\caption{Achievable throughput, normalized to $I_{DBF}$, by the proposed approach
with the strongly coherent codebook and comparisons between \textit{$I_{Pro}(Eig,M=3)$}
and its approximations.\label{fig: coherent_cb} }
\end{figure}

With the other non-orthogonal codebook whose coherence is $0.99$,
see Fig. \ref{fig: coherent_cb}, unfortunately the throughput degrades
with the increasing $M$ when $\text{SNR}<10\text{ dB}$. From (\ref{eq: vec_Z}),
it is clear that, when we select some highly correlated columns of
the codebook, that the variances of the elements of $\text{vec}(\mathbf{Z}_{E,i}[k])$
are increased leads to unreliable estimates of the effective channel,
especially in the low SNR regime. With larger $M$, it has a higher
probability of selecting these unreliable estimates.

\begin{figure}[t]
\centering{}\hspace*{-0.3cm}\includegraphics[scale=0.58]{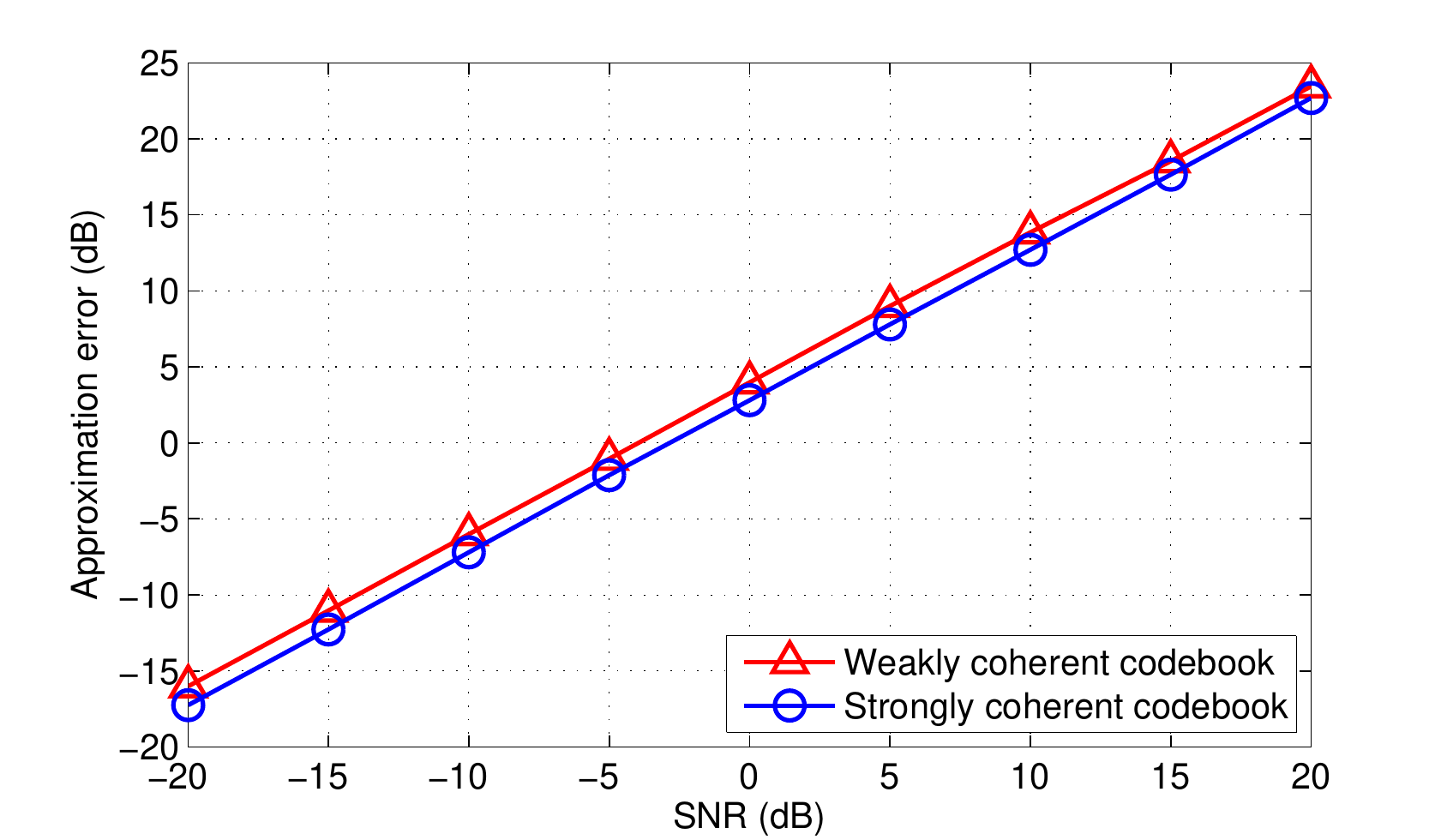}\caption{Approximation error ($\epsilon$) between \textit{$I_{Pro}(Eig,N\!F,M=3)$}
and \textit{$I_{Pro}(Fro,N\!F,M=3)$}.\label{fig: appro-error} }
\end{figure}

In Fig. \ref{fig: incoherent_cb} and Fig. \ref{fig: coherent_cb},
when $\text{SNR}>5\text{ dB}$, the gap between \textit{$I_{Pro}(Eig,M=3)$}
and \textit{$I_{Pro}(Fro,M=3)$} is obvious, but it is not clear that
either the approximation error between (\ref{eq: I_max}) and (\ref{eq: Frob_norm})
or the effective noise dominates the performance loss. To this end,
we further provide Fig. \ref{fig: appro-error} to show the approximation
error, which are calculated by the following steps. First, using \textbf{Algorithm
1} with the noise-free observations and selection criterion $f(\mathbf{H}_{E,i}[k])=\sum_{n_{s}=1}^{N_{S}}\log_{2}(1+\gamma\left[\boldsymbol{\mathbf{\Sigma}}_{E,i}^{2}[k]\right]_{n_{s},n_{s}})$
to obtain $(\hat{\mathbf{F}}_{P},\hat{\mathbf{W}}_{P})$ and the corresponding
noise-free effective channel matrix written as
\begin{equation}
\mathbf{H}_{E,\hat{i}}[k]=(\hat{\mathbf{W}}_{P}^{H}\hat{\mathbf{W}}_{P})^{-0.5}\hat{\mathbf{W}}_{P}^{H}\mathbf{H}[k]\hat{\mathbf{F}}_{P}(\hat{\mathbf{F}}_{P}^{H}\hat{\mathbf{F}}_{P})^{-0.5}.
\end{equation}
Then using $\mathbf{H}_{E,\hat{i}}[k]$ to calculate the approximation
error between (\ref{eq: I_max}) and (\ref{eq: Frob_norm}) given
by

{\small{}
\begin{equation}
\begin{aligned}\epsilon & =\frac{1}{K}\left|\text{E}\left[\sum_{k=0}^{K-1}\underset{(\ref{eq: I_max})}{\underbrace{\sum_{n_{s}=1}^{N_{S}}\log_{2}\left(1+\gamma\left[\boldsymbol{\mathbf{\Sigma}}_{E,\hat{i}}^{2}[k]\right]_{n_{s},n_{s}}\right)}}\right]\right.\\
 & \qquad\qquad\qquad\qquad\qquad\qquad\left.-\text{E}\left[\sum_{k=0}^{K-1}\underset{(\ref{eq: Frob_norm})}{\underbrace{\gamma\left\Vert \mathbf{H}_{E,\hat{i}}[k]\right\Vert _{F}^{2}}}\right]\right|\\
 & \approx\frac{\gamma}{K}\left(\frac{1}{\ln(2)}-1\right)\cdot\text{E}\left[\sum_{k=0}^{K-1}\left\Vert \mathbf{H}_{E,\hat{i}}[k]\right\Vert _{F}^{2}\right],
\end{aligned}
\label{eq: epsilon}
\end{equation}
}where the diagonal elements of $\boldsymbol{\mathbf{\Sigma}}_{E,\hat{i}}^{2}$
are the eigenvalues of $\mathbf{H}_{E,\hat{i}}\mathbf{H}_{E,\hat{i}}^{H}$.
Repeating the steps with the two non-orthogonal codebooks yields the
approximation errors shown in Fig. \ref{fig: appro-error}. The approximation
error is proportional to the SNR value (or $\gamma$). As a result,
at high SNR, the gaps between \textit{$I_{Pro}(Eig,M=3)$} and \textit{$I_{Pro}(Fro,M=3)$}
in Fig. \ref{fig: incoherent_cb} and Fig. \ref{fig: coherent_cb}
become obvious. At low SNR, because the approximation error between
(\ref{eq: I_max}) and (\ref{eq: Frob_norm}) is small, we can also
use the analysis results of (\ref{eq: Frob_norm}) in Section \ref{subsec:Orthogonal-codebooks}
and Section \ref{subsec:Non-orthogonal-Codebooks} to explain the
noise effect on $I_{Pro}(Eig,M=2,3)$ with the orthogonal and non-orthogonal
codebooks. Moreover, at low SNR, the approximation error between \textit{$I_{Pro}(Eig,M=3)$}
and \textit{$I_{Pro}(Fro,M=3)$} with the strongly coherent codebook
in Fig. \ref{fig: coherent_cb} seems larger than the approximation
error with the weakly coherent codebook in Fig. \ref{fig: incoherent_cb}.
Nevertheless, from Fig. \ref{fig: appro-error}, we can find that
without considering the noise effect on the coupling coefficients,
the approximation error with the strongly coherent codebook is even
smaller than the results with the other codebook. 

In the numerical results, we use these three different types of codebooks
to analyze the performance of the proposed method. Generally speaking,
the proposed method works well by using orthogonal and weakly coherent
codebooks.

\section{Conclusion}

This paper presents a novel strategy for the implementation of hybrid
beamforming. It shows that hybrid beamforming matrices at the transmitter
and receiver can be easily implemented based on the received coupling
coefficients so that channel estimation and singular value decomposition
for large antenna arrays are unnecessary. The idea behind this approach
is simple: efficiently evaluating the key parameters of the hybrid
beamforming gain, such as the Frobenius norm of the effective channel
or the absolute value of the determinant of the effective channel.
Since the key parameters are functions of the effective channel matrix,
which has a much smaller size typically, it is not difficult to try
a (small) set of possible alternatives to find a reasonable approximation
of the optimal hybrid beamforming matrices. The improvement achieved
by additional alternatives can be viewed as a diversity effect that
is available from multiple different pairs of array patterns. Moreover,
the effective channel matrix can be obtained from the estimated coupling
coefficients. This avoids acquiring \textit{explicit} channel estimates
and knowledge of the specific angles of propagation paths. In turns
out that \textit{implicit} channel knowledge in the sense of which
beam pairs produce the strongest coupling between the transmitter
and receiver is sufficient. Compared with hybrid beamforming methods
based on the explicit CSI, the proposed algorithm facilitates the
low-complexity hybrid beamforming implementation and should be even
robust to deviations from certain desired ideal beam patterns, which
is a weak point of all methods based on compressed sensing techniques.

\appendices{}

\section{Derivation of $\overline{\mathbf{F}}_{B,i}[k]$ and $\overline{\mathbf{W}}_{B,i}[k]$\label{sec:Derivation-of-DBF}}

In the problem (\ref{eq:constrained_DBF}), if there exists $\mathbf{W}_{B}[k]$
such that
\begin{equation}
\mathbf{W}_{B}^{H}[k]\overline{\mathbf{W}}_{P,i_{w}}^{H}\overline{\mathbf{W}}_{P,i_{w}}\mathbf{W}_{B}[k]=\mathbf{I}_{N_{S}},\label{eq: condition-2}
\end{equation}
one can define a matrix $\mathbf{Q}_{W}[k]=(\overline{\mathbf{W}}_{P,i_{w}}^{H}\overline{\mathbf{W}}_{P,i_{w}})^{0.5}\mathbf{W}_{B}[k]$,
which is equivalent to $\mathbf{W}_{B}[k]=(\overline{\mathbf{W}}_{P,i_{w}}^{H}\overline{\mathbf{W}}_{P,i_{w}})^{-0.5}\mathbf{Q}_{W}[k]$.
Replacing $\mathbf{W}_{B}[k]$ in (\ref{eq: condition-2}) by $(\overline{\mathbf{W}}_{P,i_{w}}^{H}\overline{\mathbf{W}}_{P,i_{w}})^{-0.5}\mathbf{Q}_{W}[k]$
leads to $\mathbf{Q}_{W}^{H}[k]\mathbf{Q}_{W}[k]=\mathbf{I}_{N_{S}}$
so that the columns of $\mathbf{Q}_{W}[k]$ are mutually orthogonal.
Similarly, if there exists $\mathbf{F}_{B}[k]$ that satisfies the
other power constraint at the transmitter, we can define another matrix
$\mathbf{Q}_{F}[k]=(\overline{\mathbf{F}}_{P,i_{f}}^{H}\overline{\mathbf{F}}_{P,i_{f}})^{0.5}\mathbf{F}_{B}[k]$,
which is equivalent to $\mathbf{F}_{B}[k]=(\overline{\mathbf{F}}_{P,i_{f}}^{H}\overline{\mathbf{F}}_{P,i_{f}})^{-0.5}\mathbf{Q}_{F}[k]$
\cite{Alkhateeb2016b,Sohrabi2016b}.

Given $\mathbf{W}_{B}[k]=(\overline{\mathbf{W}}_{P,i_{w}}^{H}\overline{\mathbf{W}}_{P,i_{w}})^{-0.5}\mathbf{Q}_{W}[k]$
and $\mathbf{F}_{B}[k]=(\overline{\mathbf{F}}_{P,i_{f}}^{H}\overline{\mathbf{F}}_{P,i_{f}})^{-0.5}\mathbf{Q}_{F}[k]$,
the objective function of the problem therefore becomes

\vspace{-0.3cm}{\small{}
\begin{flalign*}
 & I(\overline{\mathbf{F}}_{P,i_{f}},\overline{\mathbf{W}}_{P,i_{w}},\mathbf{F}_{B}[k],\mathbf{W}_{B}[k])\\
 & =\log_{2}\det\left(\mathbf{I}_{N_{S}}+\frac{1}{\sigma_{n}^{2}}\mathbf{Q}_{W}^{H}[k]\mathbf{H}_{E,i}[k]\mathbf{Q}_{F}[k]\mathbf{R}_{s}\mathbf{Q}_{F}^{H}[k]\mathbf{H}_{E,i}^{H}[k]\mathbf{Q}_{W}[k]\right),
\end{flalign*}
}where $\mathbf{H}_{E,i}[k]$, $i=(i_{f}-1)I_{W}+i_{w}$, is the effective
channel defined as
\[
\begin{alignedat}{1}\mathbf{H}_{E,i}[k] & \triangleq(\overline{\mathbf{W}}_{P,i_{w}}^{H}\overline{\mathbf{W}}_{P,i_{w}})^{-0.5}\overline{\mathbf{W}}_{P,i_{w}}^{H}\mathbf{H}[k]\overline{\mathbf{F}}_{P,i_{f}}(\overline{\mathbf{F}}_{P,i_{f}}^{H}\overline{\mathbf{F}}_{P,i_{f}})^{-0.5}.\end{alignedat}
\]
Let the SVD of $\mathbf{H}_{E,i}[k]$ be
\[
\mathbf{H}_{E,i}[k]\overset{\text{SVD}}{=}\mathbf{U}_{E,i}[k]\boldsymbol{\mathbf{\Sigma}}_{E,i}[k]\mathbf{V}_{E,i}^{H}[k],
\]
the throughput at subcarrier $k$ is bounded by
\begin{multline*}
I(\overline{\mathbf{F}}_{P,i_{f}},\overline{\mathbf{W}}_{P,i_{w}},\mathbf{F}_{B}[k],\mathbf{W}_{B}[k])\\
\leq\sum_{n_{s}=1}^{N_{S}}\log_{2}\left(1+\frac{1}{\sigma_{n}^{2}}\left[\boldsymbol{\mathbf{\Sigma}}_{E,i}^{2}[k]\right]_{n_{s},n_{s}}\left[\mathbf{R}_{s}\right]_{n_{s},n_{s}}\right)
\end{multline*}
with equality if $\mathbf{Q}_{W}[k]=\left[\mathbf{U}_{E,i}[k]\right]_{:,1:N_{S}}$,
where the columns of $\mathbf{Q}_{W}[k]$ are mutually orthogonal
as required, and $\mathbf{Q}_{F}[k]=\left[\mathbf{V}_{E,i}[k]\right]_{:,1:N_{S}}$,
which satisfies the condition $\text{tr}(\overline{\mathbf{F}}_{P,i_{f}}\mathbf{F}_{B}[k]\mathbf{R}_{s}\mathbf{F}_{B}^{H}[k]\overline{\mathbf{F}}_{P,i_{f}}^{H})=\text{tr}(\mathbf{R}_{s})$
when $\mathbf{F}_{B}[k]=(\overline{\mathbf{F}}_{P,i_{f}}^{H}\overline{\mathbf{F}}_{P,i_{f}})^{-0.5}\mathbf{Q}_{F}[k]$.
As a result, the solution to the maximization problem is given by

\begin{align*}
\overline{\mathbf{F}}_{B,i}[k] & =(\overline{\mathbf{F}}_{P,i_{f}}^{H}\overline{\mathbf{F}}_{P,i_{f}})^{-0.5}\left[\mathbf{V}_{E,i}[k]\right]_{:,1:N_{S}},\\
\overline{\mathbf{W}}_{B,i}[k] & =(\overline{\mathbf{W}}_{P,i_{w}}^{H}\overline{\mathbf{W}}_{P,i_{w}})^{-0.5}\left[\mathbf{U}_{E,i}[k]\right]_{:,1:N_{S}}.
\end{align*}

\begin{figure*}[t]
{\small{}\vspace{-0.3cm}}
\begin{equation}
\begin{alignedat}{1}\mathbf{R}_{Z_{E,i}} & \triangleq\text{E}\left[\text{vec}(\mathbf{Z}_{E,i}[k])\text{vec}(\mathbf{Z}_{E,i}[k])^{H}\right]\\
 & =\text{E}\left[\left((\overline{\mathbf{F}}_{P,i_{f}}^{T}\overline{\mathbf{F}}_{P,i_{f}}^{*})^{-0.5}\otimes(\overline{\mathbf{W}}_{P,i_{w}}^{H}\overline{\mathbf{W}}_{P,i_{w}})^{-0.5}\right)\text{vec}(\mathbf{Z}[k])\text{vec}(\mathbf{Z}[k])^{H}\left((\overline{\mathbf{F}}_{P,i_{f}}^{T}\overline{\mathbf{F}}_{P,i_{f}}^{*})^{-0.5}\otimes(\overline{\mathbf{W}}_{P,i_{w}}^{H}\overline{\mathbf{W}}_{P,i_{w}})^{-0.5}\right)^{H}\right]\\
 & =\sigma_{n}^{2}\left((\overline{\mathbf{F}}_{P,i_{f}}^{T}\overline{\mathbf{F}}_{P,i_{f}}^{*})^{-1}\otimes(\overline{\mathbf{W}}_{P,i_{w}}^{H}\overline{\mathbf{W}}_{P,i_{w}})^{-1}\right)
\end{alignedat}
\label{eq: R_Z_V}
\end{equation}
{\footnotesize{}\hrulefill}
\end{figure*}

\begin{figure*}[t]
{\small{}\vspace{-0.3cm}}
\begin{equation}
\begin{alignedat}{1}\text{Var}(U) & =\text{E}\left[U^{2}\right]-\text{E}\left[U\right]^{2}=\text{E}\left[\left\Vert \mathbf{Z}_{E,i}[k]\right\Vert _{F}^{4}\right]-\text{E}\left[U\right]^{2}\\
 & =\text{E}\underset{\qquad\boldsymbol{\Psi}\in\mathbb{C}^{N_{RF}^{4}\times N_{RF}^{4}}}{\left[\text{tr}\left(\underbrace{\left(\left((\overline{\mathbf{F}}_{P,i_{f}}^{T}\overline{\mathbf{F}}_{P,i_{f}}^{*})^{-1}\otimes(\overline{\mathbf{W}}_{P,i_{w}}^{H}\overline{\mathbf{W}}_{P,i_{w}})^{-1}\right)\otimes\left((\overline{\mathbf{F}}_{P,i_{f}}^{T}\overline{\mathbf{F}}_{P,i_{f}}^{*})^{-1}\otimes(\overline{\mathbf{W}}_{P,i_{w}}^{H}\overline{\mathbf{W}}_{P,i_{w}})^{-1}\right)\right)}\right.\right.}\\
 & \quad\,\cdot\left.\left.\left(\left(\text{vec}(\mathbf{Z}[k])\text{vec}(\mathbf{Z}[k])^{H}\right)\otimes\left(\text{vec}(\mathbf{Z}[k])\text{vec}(\mathbf{Z}[k])^{H}\right)\right)\right)\right]-\text{E}\left[U\right]^{2}\\
 & =\text{tr}\left(\boldsymbol{\Psi}\cdot\text{E}\left[\left(\text{vec}(\mathbf{Z}[k])\text{vec}(\mathbf{Z}[k])^{H}\right)\otimes\left(\text{vec}(\mathbf{Z}[k])\text{vec}(\mathbf{Z}[k])^{H}\right)\right]\right)-\text{E}\left[U\right]^{2}\\
 & =\text{tr}\left(\boldsymbol{\Psi}\cdot\text{E}[\underset{\mathbf{z}_{V}[k]\in\mathbb{C}^{N_{RF}^{4}\times1}}{\underbrace{\left(\text{vec}(\mathbf{Z}[k])\otimes\text{vec}(\mathbf{Z}[k])\right)}}\underset{\mathbf{z}_{V}^{H}[k]}{\underbrace{\left(\text{vec}(\mathbf{Z}[k])\otimes\text{vec}(\mathbf{Z}[k])\right)^{H}}}]\right)-\text{E}\left[U\right]^{2}\\
 & =\text{tr}\left(\boldsymbol{\Psi}\cdot\mathbf{R}_{z_{V}}\right)-\text{E}\left[U\right]^{2}
\end{alignedat}
\label{eq: Var(U)}
\end{equation}
{\footnotesize{}\hrulefill}
\end{figure*}

\section{Derivation of the covariance matrix of $\text{vec}(\mathbf{Z}_{E,i}[k])$\label{sec:Derivation-of-the}}

Let us repeat (\ref{eq: H_E_hat}) that $\mathbf{Z}_{E,i}[k]=(\overline{\mathbf{W}}_{P,i_{w}}^{H}\overline{\mathbf{W}}_{P,i_{w}})^{-0.5}\mathbf{Z}[k](\overline{\mathbf{F}}_{P,i_{f}}^{H}\overline{\mathbf{F}}_{P,i_{f}})^{-0.5},$
where the elements of $\mathbf{Z}[k]$ have the same normal distribution
with mean zero and variance $\sigma_{n}^{2}$. To find the covariance
between the elements of $\mathbf{Z}_{E,i}[k]$, we vectorize $\mathbf{Z}_{E,i}[k]$
as

\[
\begin{alignedat}{1} & \text{vec}(\mathbf{Z}_{E,i}[k])\\
 & =\left(\left((\overline{\mathbf{F}}_{P,i_{f}}^{H}\overline{\mathbf{F}}_{P,i_{f}})^{-0.5}\right)^{T}\otimes(\overline{\mathbf{W}}_{P,i_{w}}^{H}\overline{\mathbf{W}}_{P,i_{w}})^{-0.5}\right)\text{vec}(\mathbf{Z}[k])\\
 & =\left((\overline{\mathbf{F}}_{P,i_{f}}^{T}\overline{\mathbf{F}}_{P,i_{f}}^{*})^{-0.5}\otimes(\overline{\mathbf{W}}_{P,i_{w}}^{H}\overline{\mathbf{W}}_{P,i_{w}})^{-0.5}\right)\text{vec}(\mathbf{Z}[k]),
\end{alignedat}
\]
and the covariance matrix of $\text{vec}(\mathbf{Z}_{E,i}[k])$ is
given by (\ref{eq: R_Z_V}).

\section{Derivation of $\text{E}\left[U\right]$ and $\text{Var}(U)$ for
non-orthogonal codebooks\label{sec:Derivation-of-}}

By the definition of $U$ in (\ref{eq: U}), one has the mean and
variance of $U$ given by
\[
\begin{alignedat}{1}\text{E}\left[U\right] & =\text{E}\left[\left\Vert \mathbf{Z}_{E,i}[k]\right\Vert _{F}^{2}\right]\\
 & =\text{E}\left[\text{tr}\left(\text{vec}\left(\mathbf{Z}_{E,i}[k]\right)\text{vec}\left(\mathbf{Z}_{E,i}[k]\right)^{H}\right)\right]\\
 & =\sigma_{n}^{2}\,\text{tr}\left((\overline{\mathbf{F}}_{P,i_{f}}^{T}\overline{\mathbf{F}}_{P,i_{f}}^{*})^{-1}\right)\text{tr}\left((\overline{\mathbf{W}}_{P,i_{w}}^{H}\overline{\mathbf{W}}_{P,i_{w}})^{-1}\right)
\end{alignedat}
\]
and $\text{Var}(U)$ in (\ref{eq: Var(U)}).

\bibliographystyle{IEEEtran}
\bibliography{reference,IEEEabrv}

\end{document}